\documentclass{aa}

\usepackage{graphicx}

\def\et{\rm et al. }
\def\mum{$\mu$m}
\def\ha{H$\alpha$}
\def\hb{H$\beta$}
\def\pa{Pa$\alpha$}
\def\bg{Br$\gamma$}

\begin{document}

 \title{NIR Spectroscopy of Luminous Infrared Galaxies and
 the Hydrogen Recombination Photon Deficit
\thanks{based on observations collected at the European Southern Observatory,
Chile, ESO No. 67.A-0593, 71.A-0707.}}

   \author{J.R. Vald\'es \inst{1,2}
          \and
           S. Berta\inst{3}
           \and
           A. Bressan\inst{1}
          \and
           A. Franceschini\inst{3}
      \and
       D. Rigopoulou\inst{4}
          \and
           G. Rodighiero \inst{3}
         }
   \authorrunning{J.R. Vald\'es et al.}
   \titlerunning{NIR Spectroscopy of LIRGs}

   \offprints{A.Bressan}

   \institute{INAF, Osservatorio Astronomico di Padova, vicolo
              dell'Osservatorio 5, 35122 Padova, Italy \\
              \email{valdes@pd.astro.it,bressan@pd.astro.it}
         \and
          Instituto Nacional de Astrof\'{\i}sica, Optica y Electr\'onica,
              Apdos. Postales 51 y 216, C.P. 72000 Puebla, Pue., M\'exico\\
              \email{jvaldes@inaoep.mx}
         \and
             Dipartimento di Astronomia, vicolo
             dell'Osservatorio 5, 35122 Padova, Italy\\
         \email{berta@pd.astro.it,rodighiero@pd.astro.it,franceschini@pd.astro.it}
     \and
     Department of Physics (Astrophysics), University of Oxford, Denys Wilkinson
     Building, 1 Keble Road, Oxford OX1 3RH, UK\\
                  \email{dar@astro.ox.ac.uk}
             }

   \date{Received date; accepted date}

\abstract{We report on near-infrared medium-resolution spectroscopy
of a sample of luminous  and ultra-luminous infrared galaxies (LIRGs-ULIRGs),
carried out with SOFI at the ESO 3.5m New Technology Telescope. 
Because of wavelength dependence of the attenuation,
the detection of
the \pa\ or \bg\ line in the Ks band should provide relevant constraints on 
SFR and  the contribution of an AGN.

We find, however, that the intensities of the \pa\ and \bg\ lines, even corrected for
slit losses, are on average only 10\% and 40\%, respectively, of that expected from a
normal starburst of similar bolometric luminosity. The corresponding star formation
rates, after correcting for the attenuation derived from the NIR-optical emission line
ratios, are 14\% and 60\% of that expected if the far 
infrared luminosity were entirely powered by the starburst.

This confirms the existence of a recombination photon deficit,
particularly in the case of the \pa\ line,
already found in the \bg\ 
line in other infrared galaxies of similar luminosity.

In discussing the possible causes of the discrepancy,
we find unlikely that it is due to the presence 
of an AGN, though  two objects show
evidence of broadening of the \pa\ line and of the presence of coronal line 
emission.
In fact, from our own observations and data collected from the literature
we argue that the studied galaxies appear to be predominantly powered by a nuclear
starburst.

Two scenarios compatible with the present data
are that either there exists a highly attenuated
nuclear star forming region, 
and/or that a significant fraction ($\simeq$80\% ) of the ionizing photons
are absorbed by dust within the HII regions.
We suggest that  observations in the Br$\alpha$ spectral region could constitute
a powerful tool to disentangle these two possibilities.

}

   \maketitle

\section{Introduction}
\label{intro}

The advent of the Infrared Space Observatory (ISO) combined with the availability of new
ground based facilities such as SCUBA on the JCMT have resulted in the discovery of numerous
distant galaxies with enhanced infrared (IR) emission (e.g. Elbaz \et \cite{elb}, Smail
\et \cite{sma}, Barger \et \cite{bar}, Gear \et \cite{gear2000}, Franceschini \et \cite{fran01},
Smail \et \cite{smaetal2003}, Ivison \et \cite{ivis2002}).

It is not yet clear what mechanisms power the bolometric luminosity of these luminous 
(LIRGs) and ultra (ULIRGs) luminous infrared galaxies, with bolometric luminosities
exceeding 10$^{11}L_\odot$ and 10$^{12}L_\odot$, respectively.

Perhaps, the tightest constraint on the nature
of the infrared luminosity is that they mostly obey the FIR/Radio correlation of normal
star forming galaxies (Sanders \& Mirabel \cite{sand1}).
However the presence of broad emission line components, a certain degree of
polarization, power-law near infrared colours, warm far-infrared (FIR) spectra, high
radio brightness temperature in the milli-arcsecond central structure and, finally,
hard X-ray emission are often detected, suggesting that a
significant contribution to the luminosity may come from mass accretion onto a
compact central object.

These galaxies are often found to be interacting systems, indicating that the two
phenomena may be triggered at once by dynamical interaction
(e.g. Rigopoulou \et \cite{rig1}).

Optical studies have been performed in order to establish the nature of these
objects (e.g. Veilleux \et \cite{veil99}), but it is difficult to disentangle between
AGN and star formation processes, because the bulk of their luminosity
is strongly attenuated by dust and the re-emitted FIR spectrum keeps little
memory of the primary power source.

Even in objects without clear signatures of the presence of AGN, there are
significant discrepancies between  different observables. For example Poggianti,
Bressan \& Franceschini (\cite{pogg}), in analyzing a sample of 
Infrared Galaxies with log(L$_{IR}$/L$_{\odot}$)$\geq$11.5, concluded that
the Star Formation Rate (SFR) detected from optical diagnostic tools (even
corrected for attenuation) amounts to about 1/3 of that inferred from the FIR.
However, if such a large fraction of stellar activity is hidden by dust, one may
wonder whether also the nuclear activity is invisible at optical wavelengths.

A natural way to avoid strong dust obscuration is to look in the near infrared
where, being the attenuation only a fraction ($\simeq$1/7) of that in the
optical, one may hope to obtain an unbiased estimate of the SFR and at the same
time to search for the possible presence of an obscured AGN by means of
broadening of permitted emission line and/or presence of coronal lines.

With these purposes we begun a medium resolution NIR spectroscopic study with
SOFI at the ESO 3.5m New Technology Telescope, of a sample of ULIRGs selected
from Genzel \et (\cite{gen}) and Rigopoulou \et (\cite{rig1}) samples. In this paper we
present and discuss these new observations.

The plan of the paper is as follows. In Sections \ref{obs} we
present the observations and data reduction, with an analysis of the emission
line properties of target galaxies presented in subsections \ref{elprop} and \ref{agnsig}.
All the process to calculate the Star Formation rates are discussed in Section \ref{sfrates}.
In particular, the technique we adopted to calculate the slit losses and to evaluate the
extinction are presented in subsections \ref{los} and \ref{att}.
In Section \ref{disc} we show that, even correcting for extinction
and losses from the slit, there is a large \pa/IR deficit, and we discuss its
possible origin. Our conclusions are summarized in Section \ref{conc}. We assume
H$_0$=65 [km/s/Mpc], $\Omega_M$=0.3, $\Omega_\Lambda$=0.7 throughout.

\begin{figure*}[!ht]
\resizebox{\hsize}{!}{\includegraphics{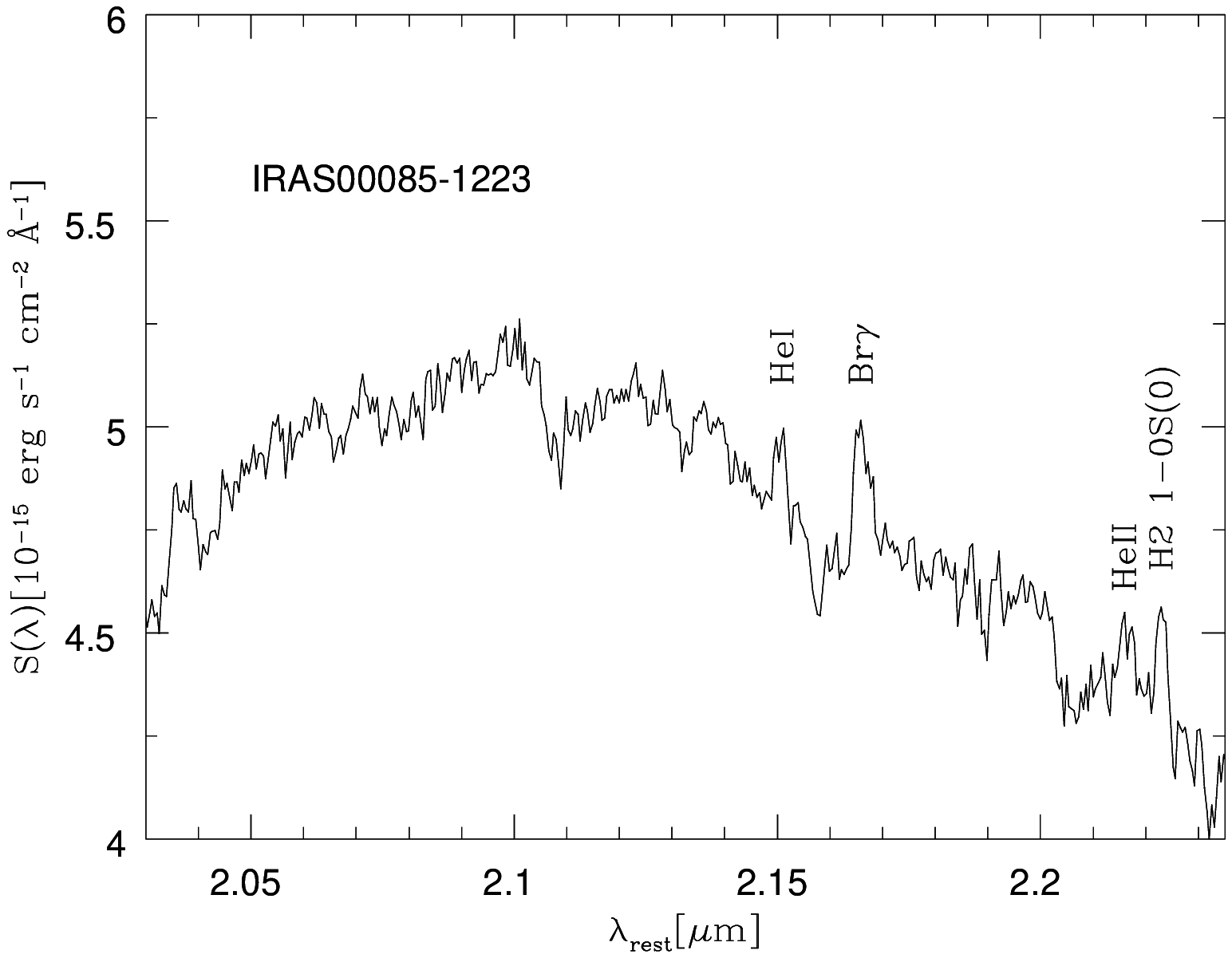}
                      \includegraphics{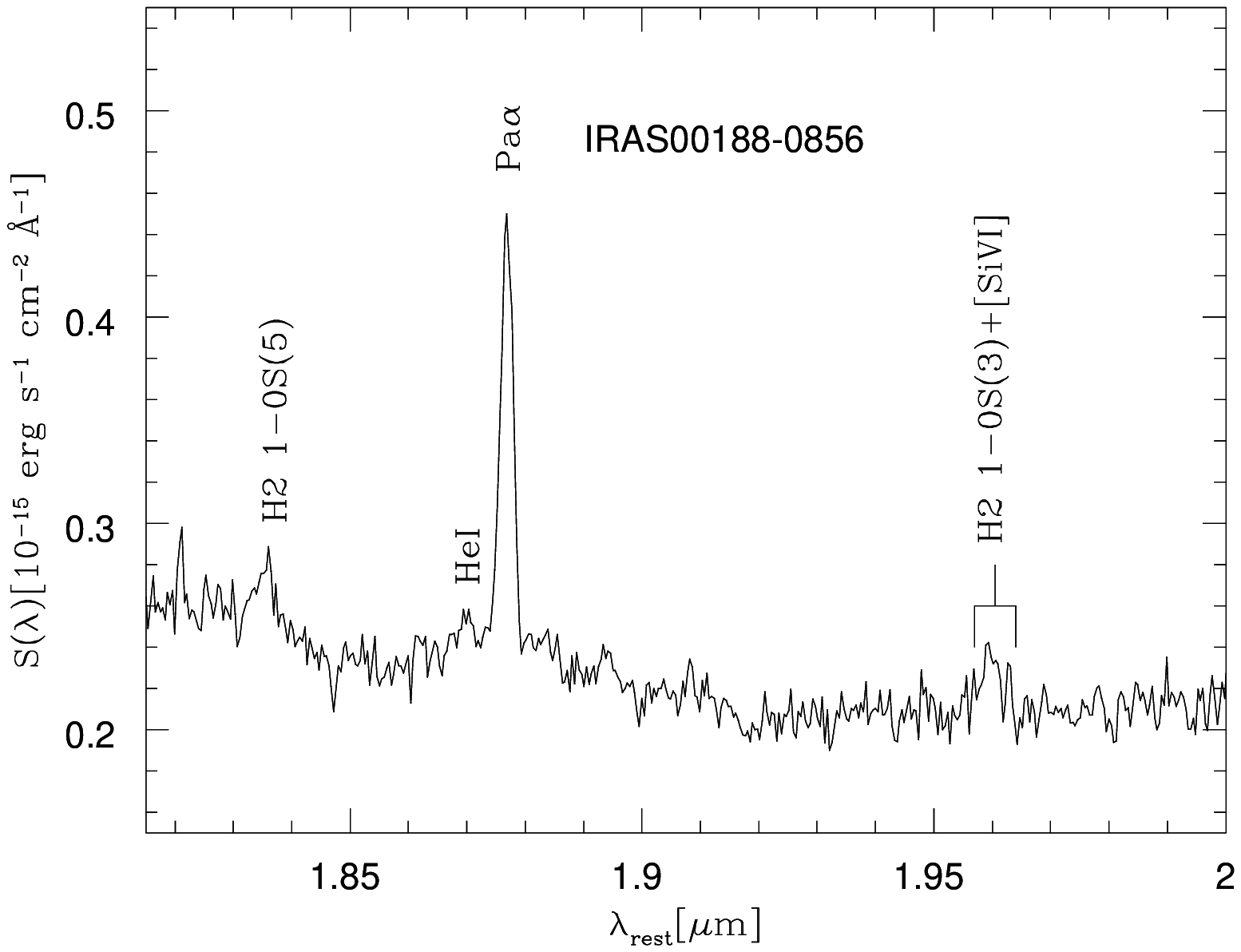}
              \includegraphics{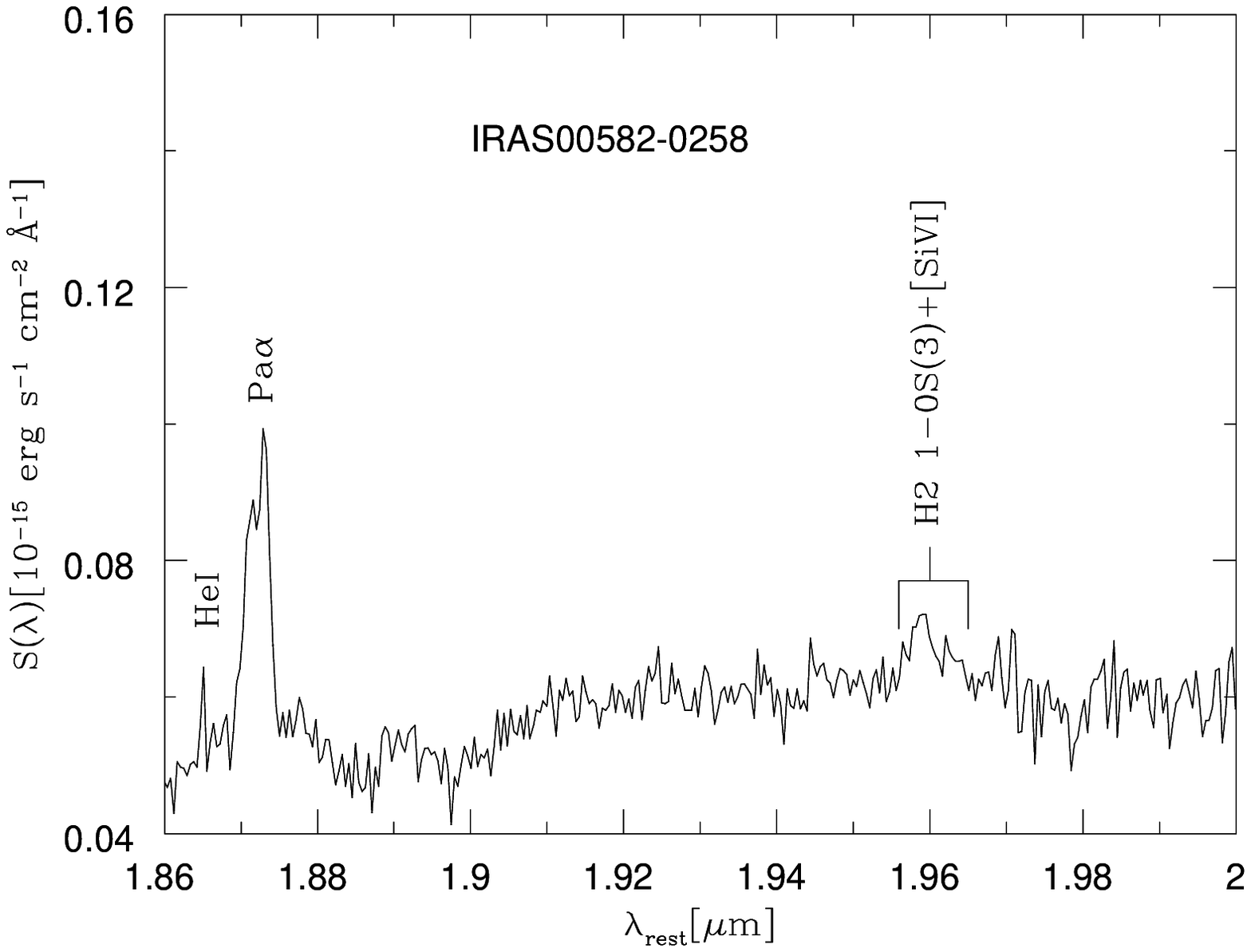}}
\resizebox{\hsize}{!}{\includegraphics{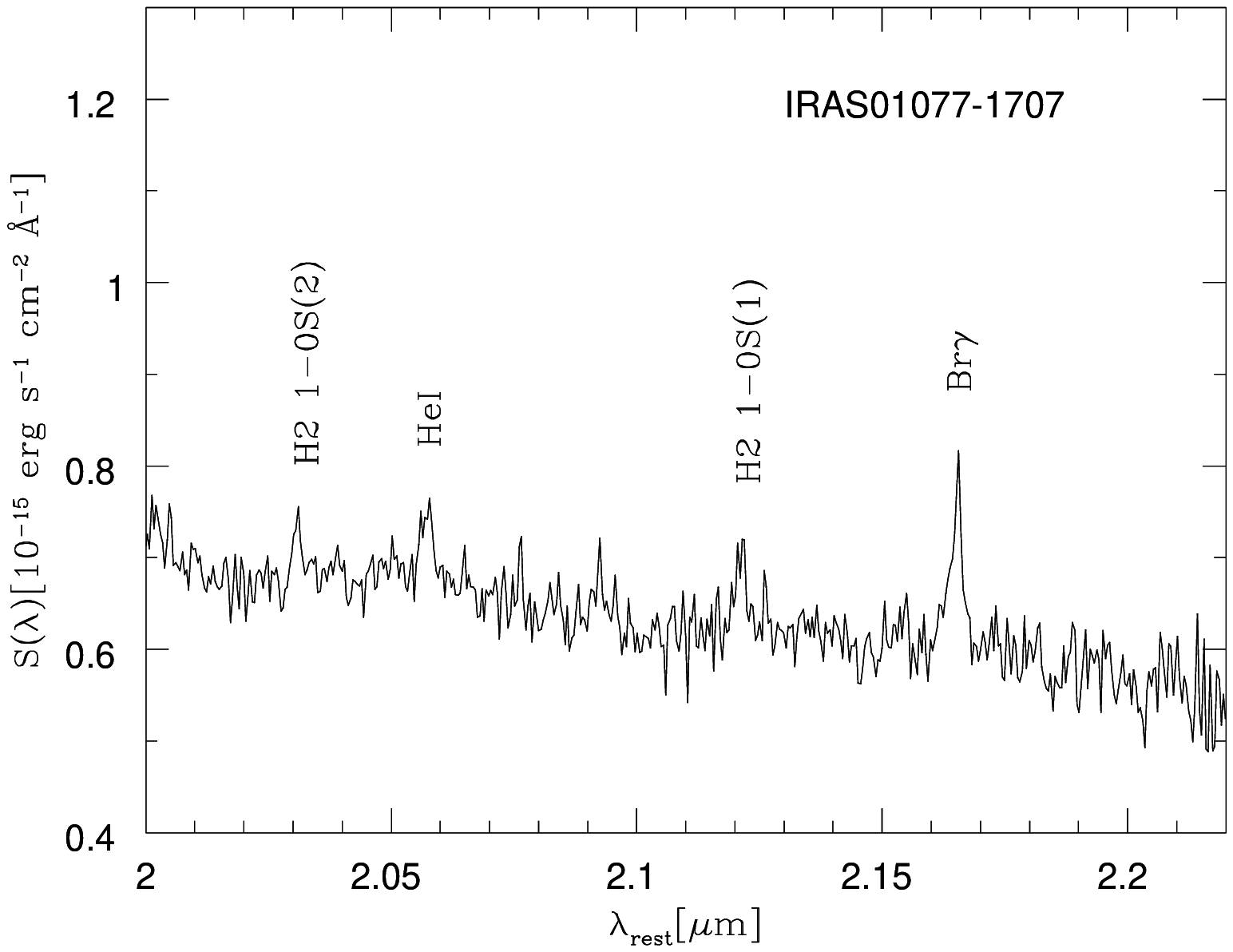}
                      \includegraphics{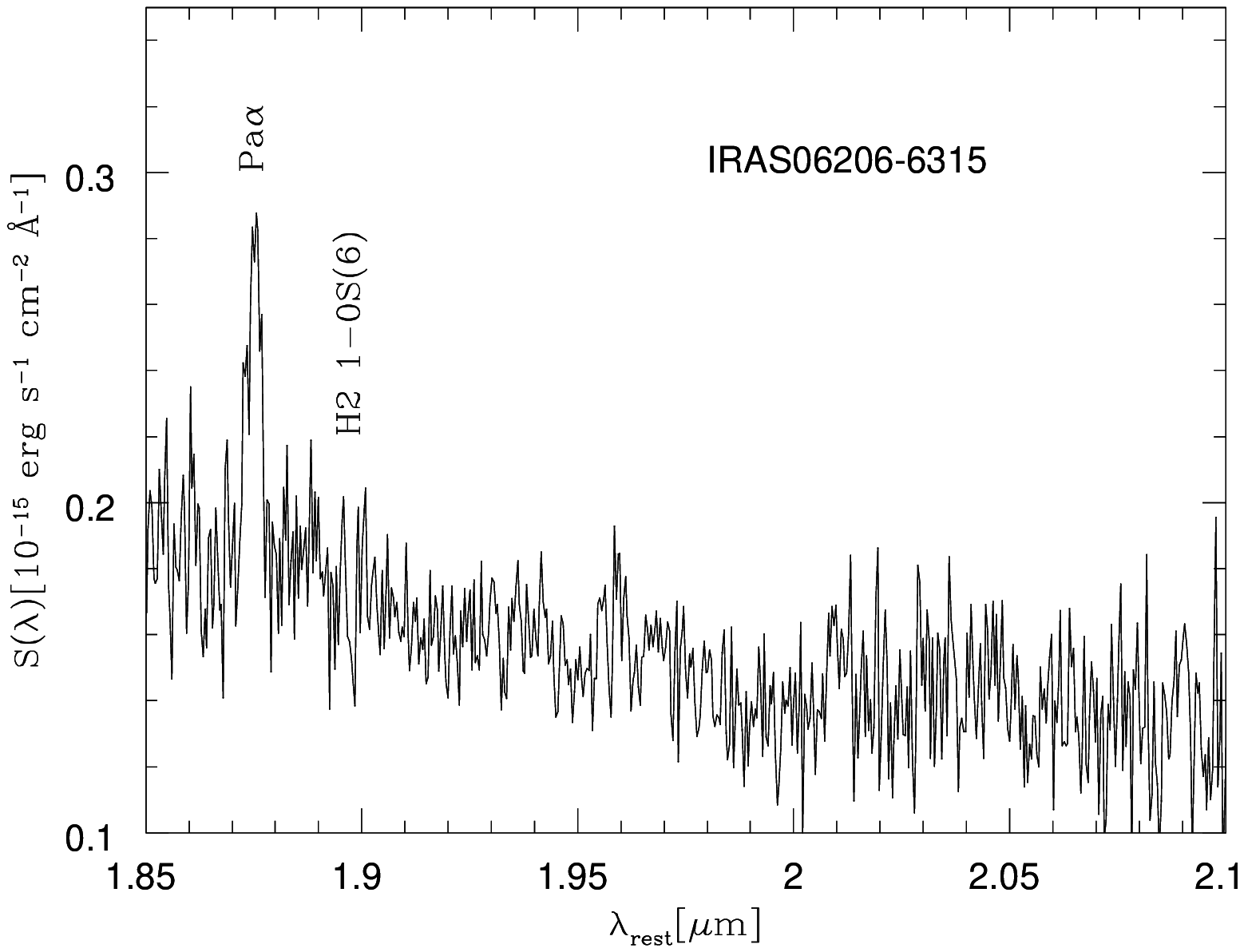}
                     \includegraphics{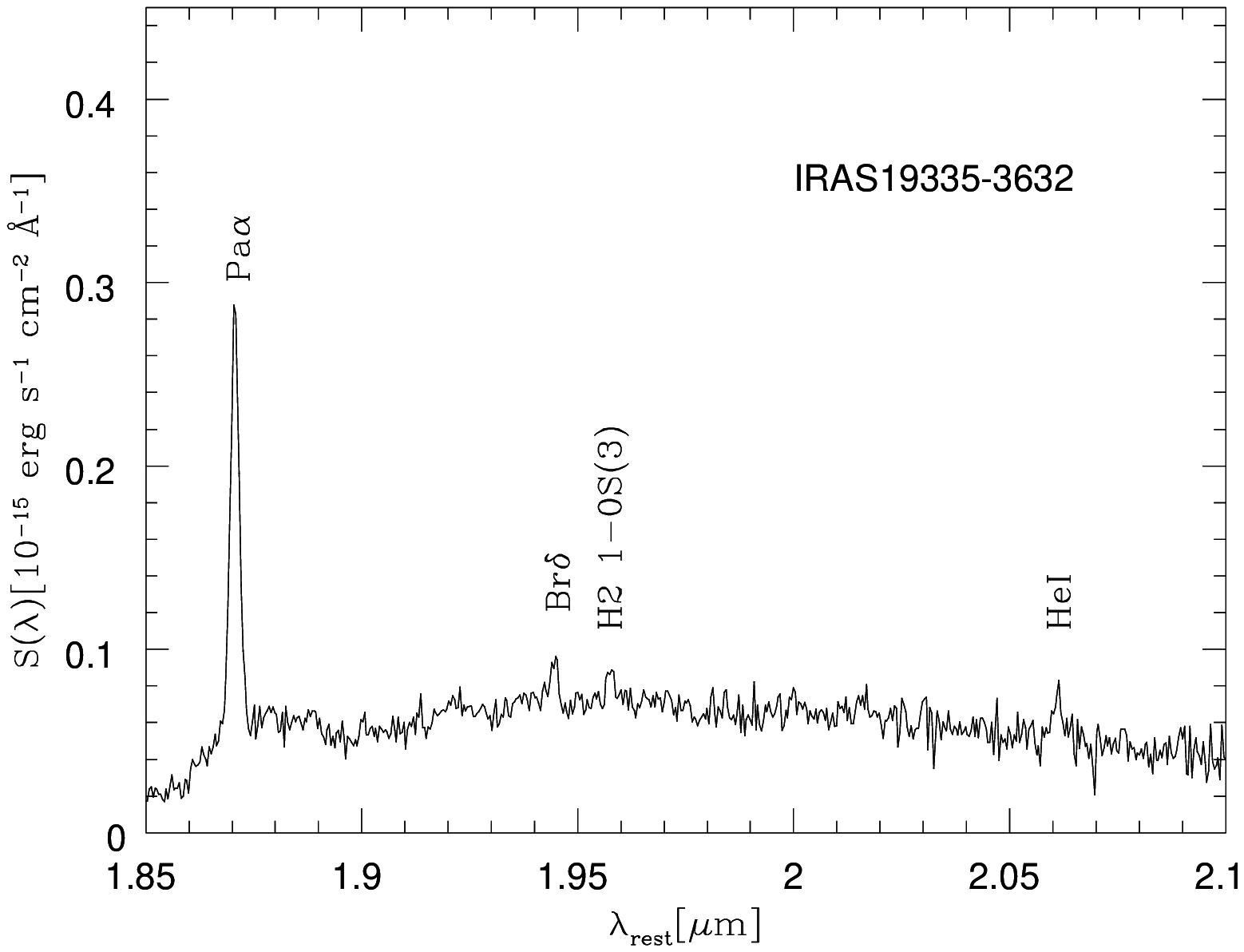}}   \resizebox{\hsize}{!}{\includegraphics{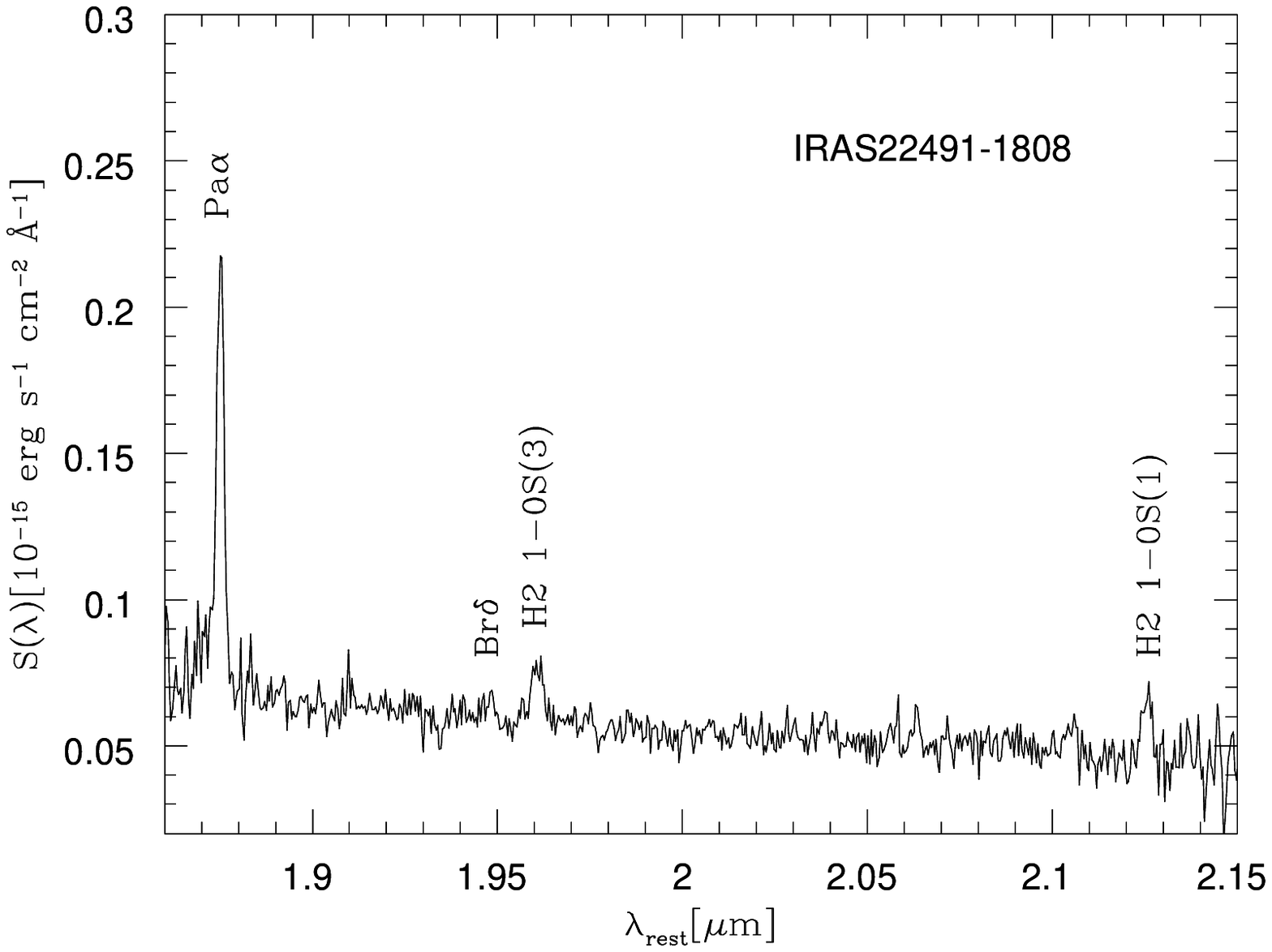}
                      \includegraphics{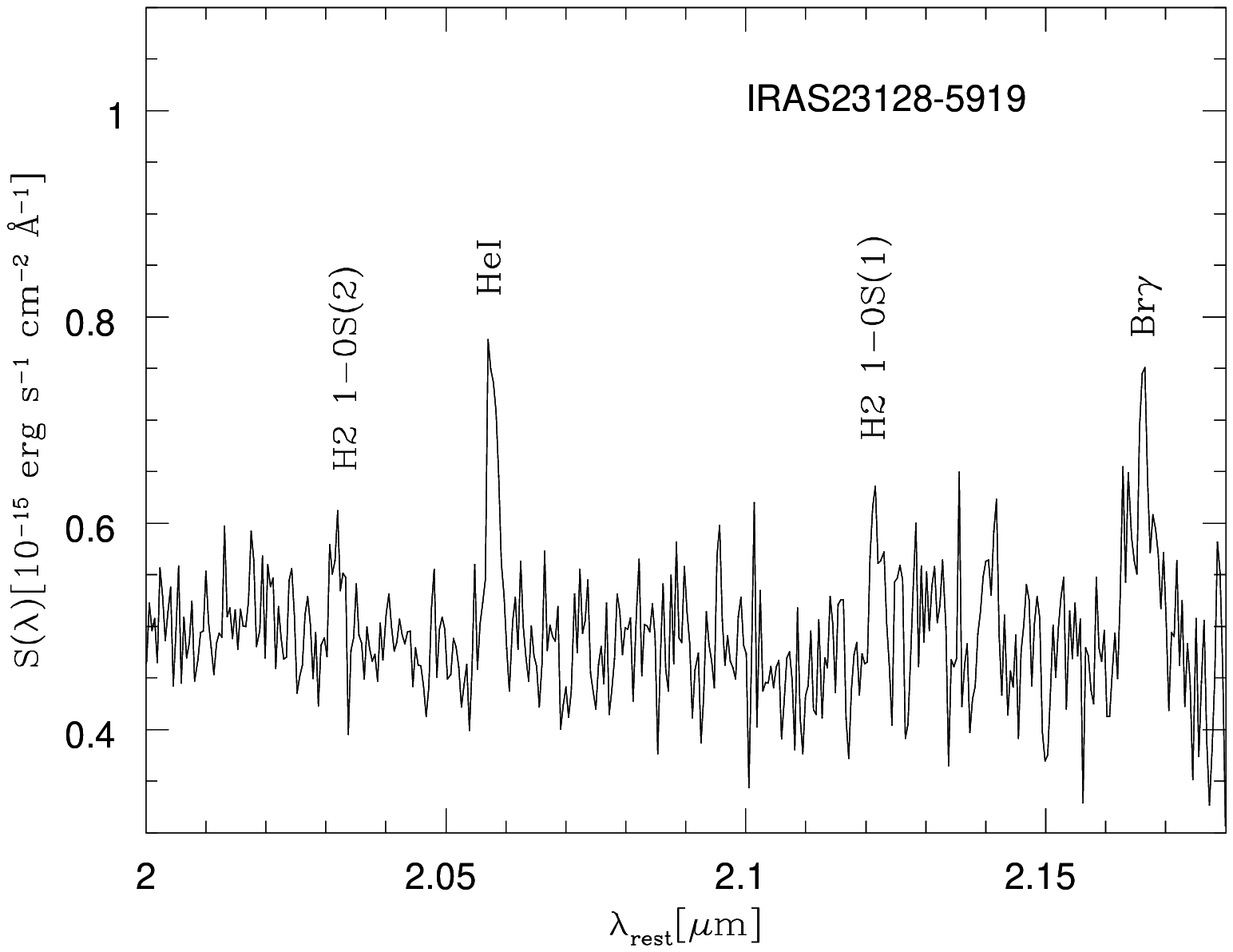}
              \includegraphics{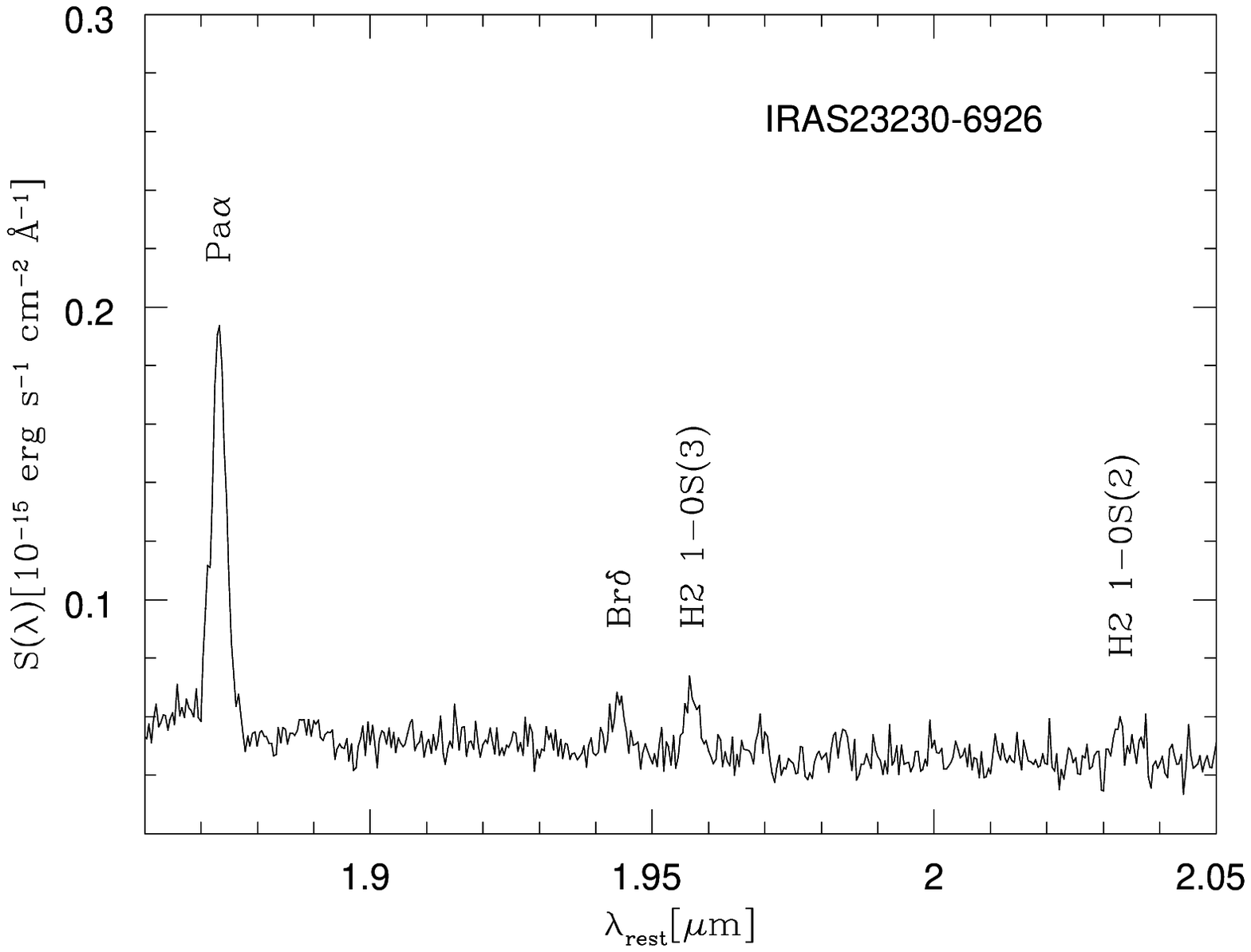}}
\begin{center}
\resizebox{!}{!}{\includegraphics[height=0.255\textwidth]
                                  {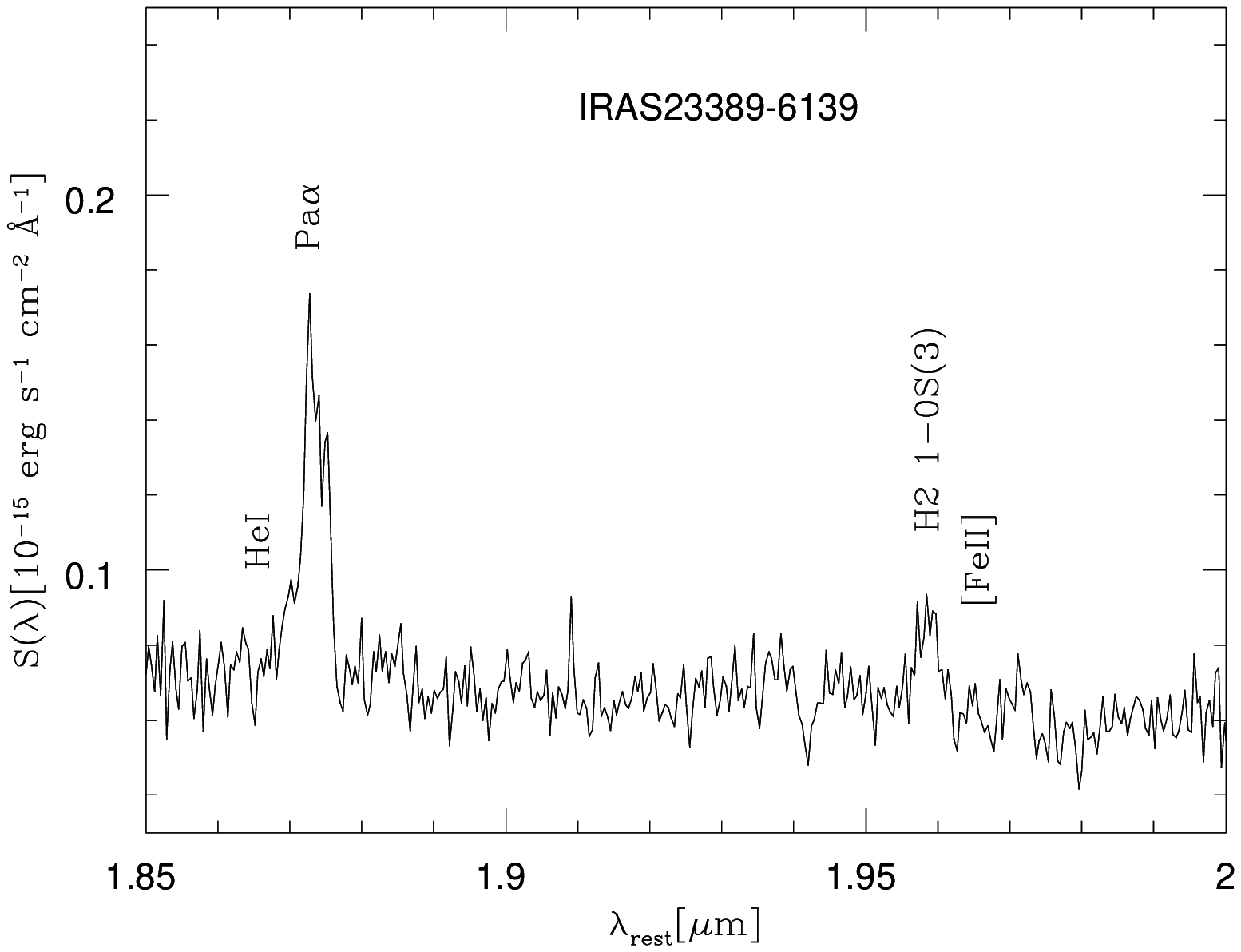}}
\end{center}
\caption{\footnotesize{Spectra of the sample galaxies showing emission in only one
nucleus. No slit-losses, nor extinction
correction have been applied. The identified emission features are highlighted. Fluxes
are in units of $[10^{-15}$ergs$^{-1}$cm$^{-2}$\AA$^{-1}]$}.}
\label{spec1}
\end{figure*}

\begin{figure*}[!ht]
\resizebox{\hsize}{!}{\includegraphics{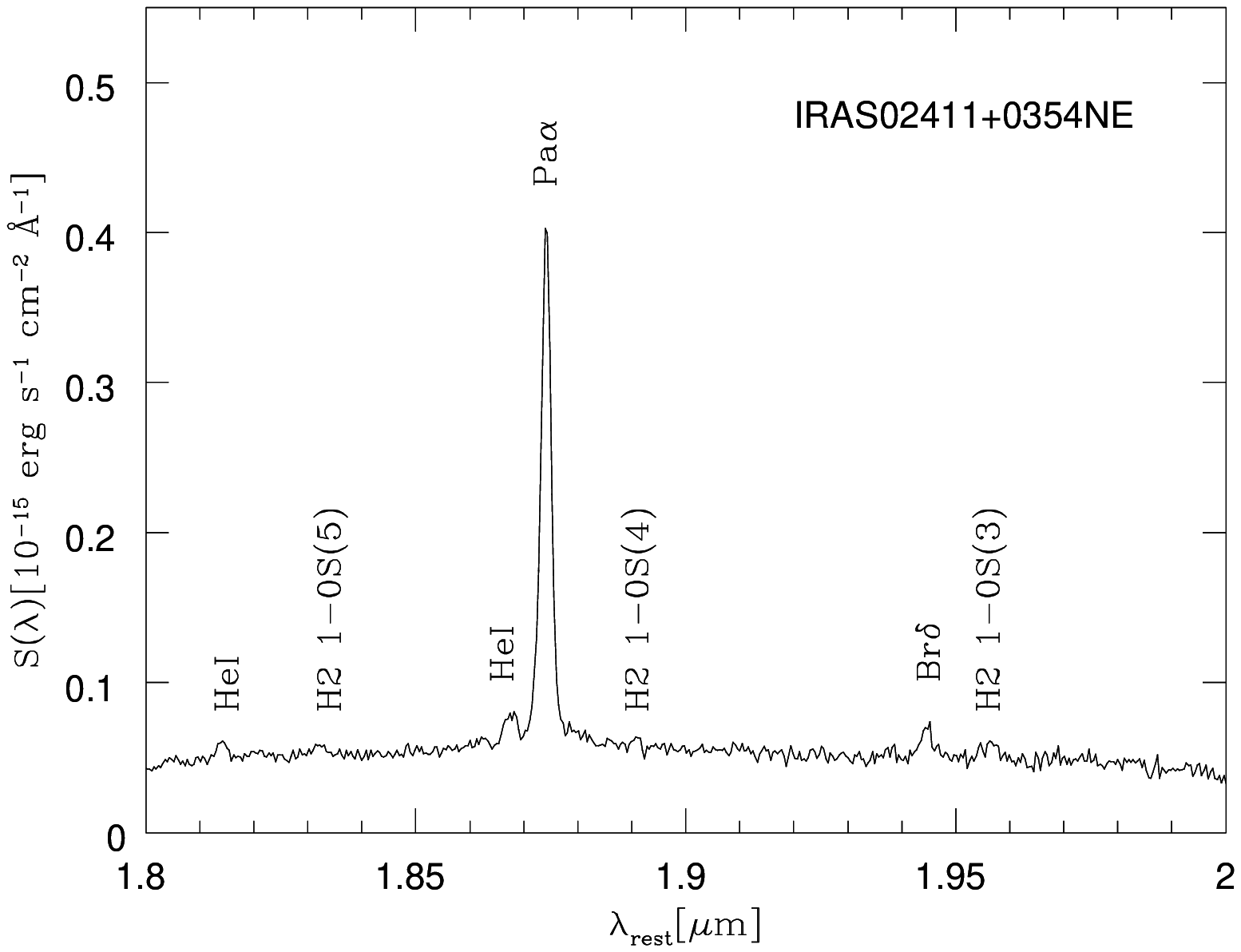}
                      \includegraphics{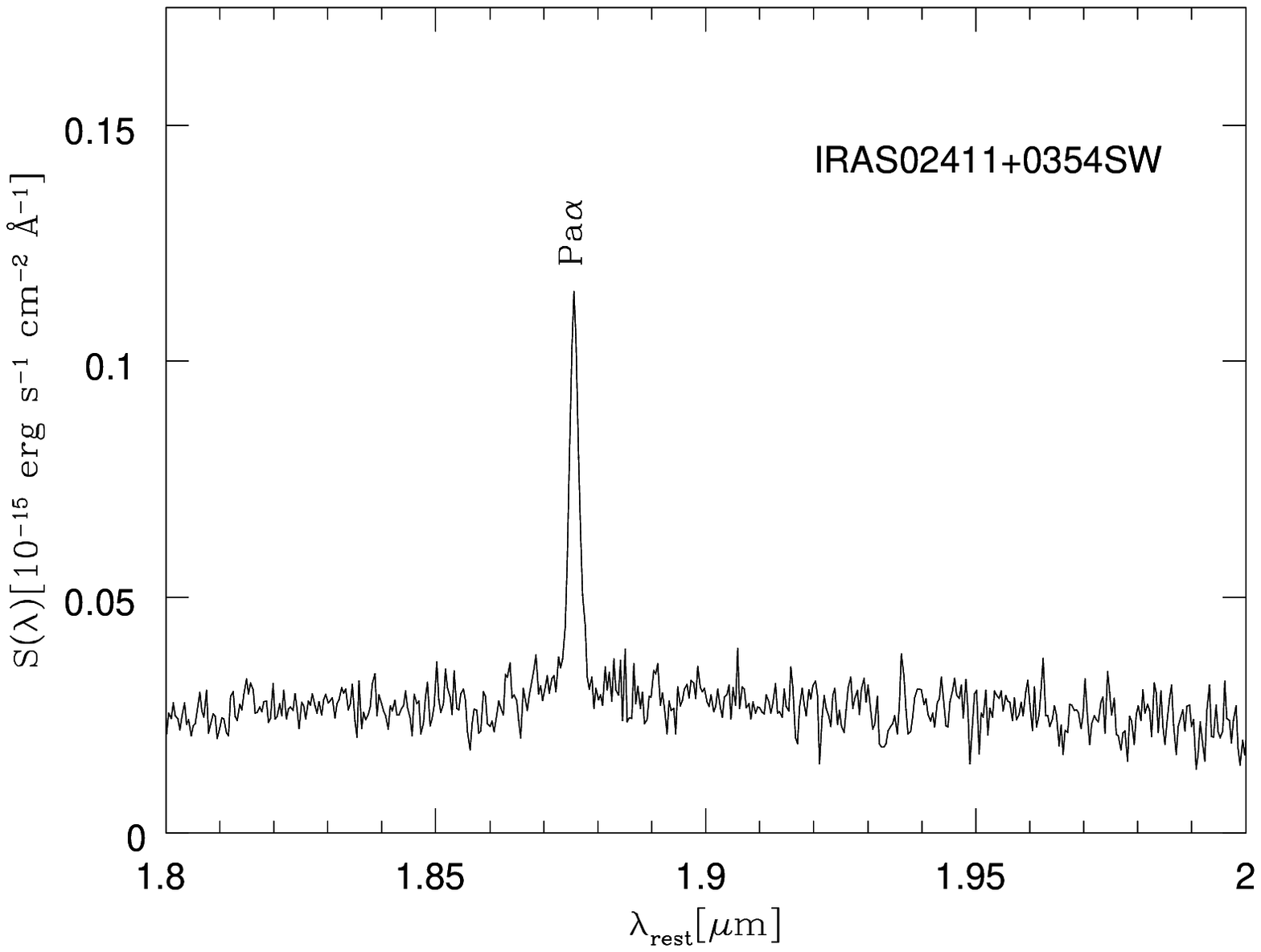}
              \includegraphics{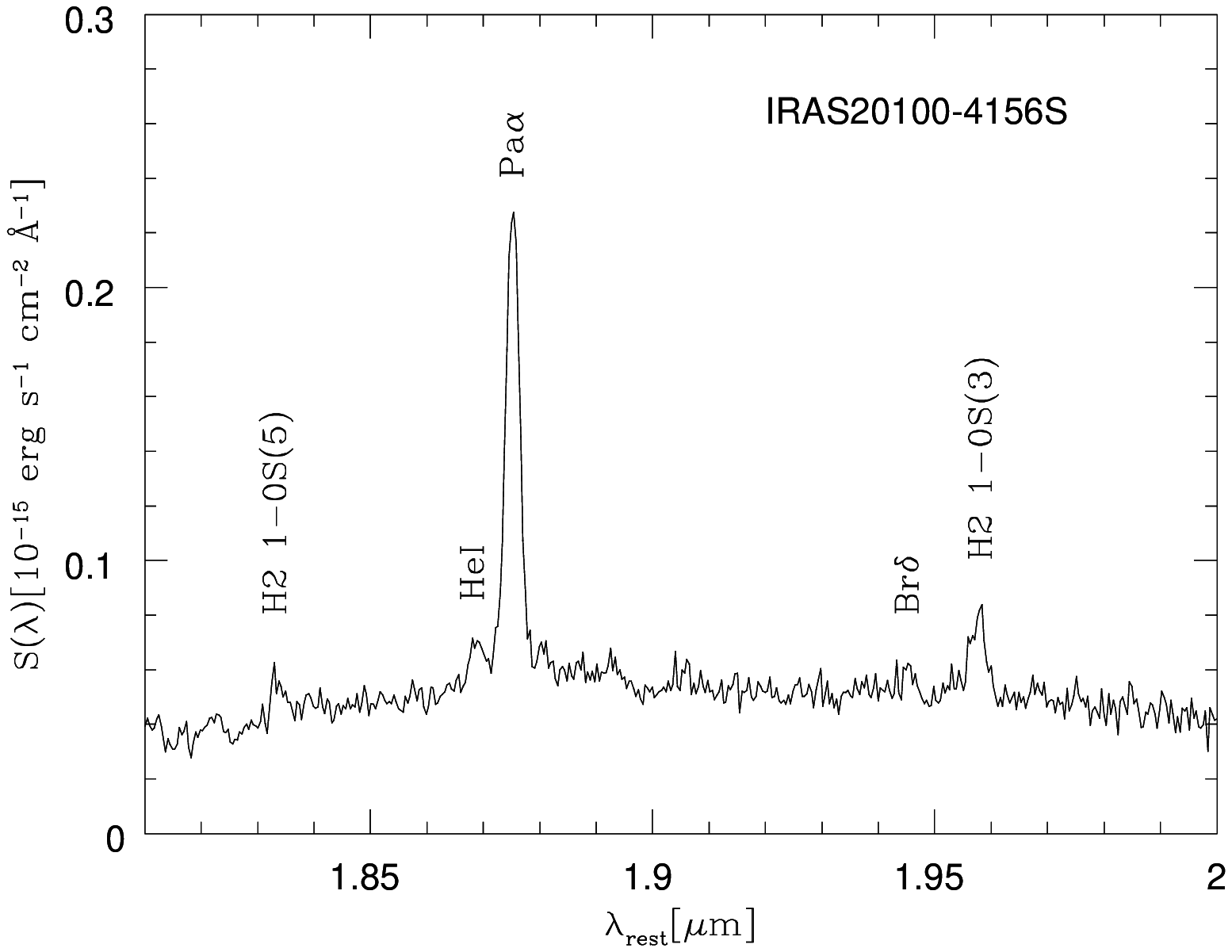}}
\resizebox{\hsize}{!}{\includegraphics{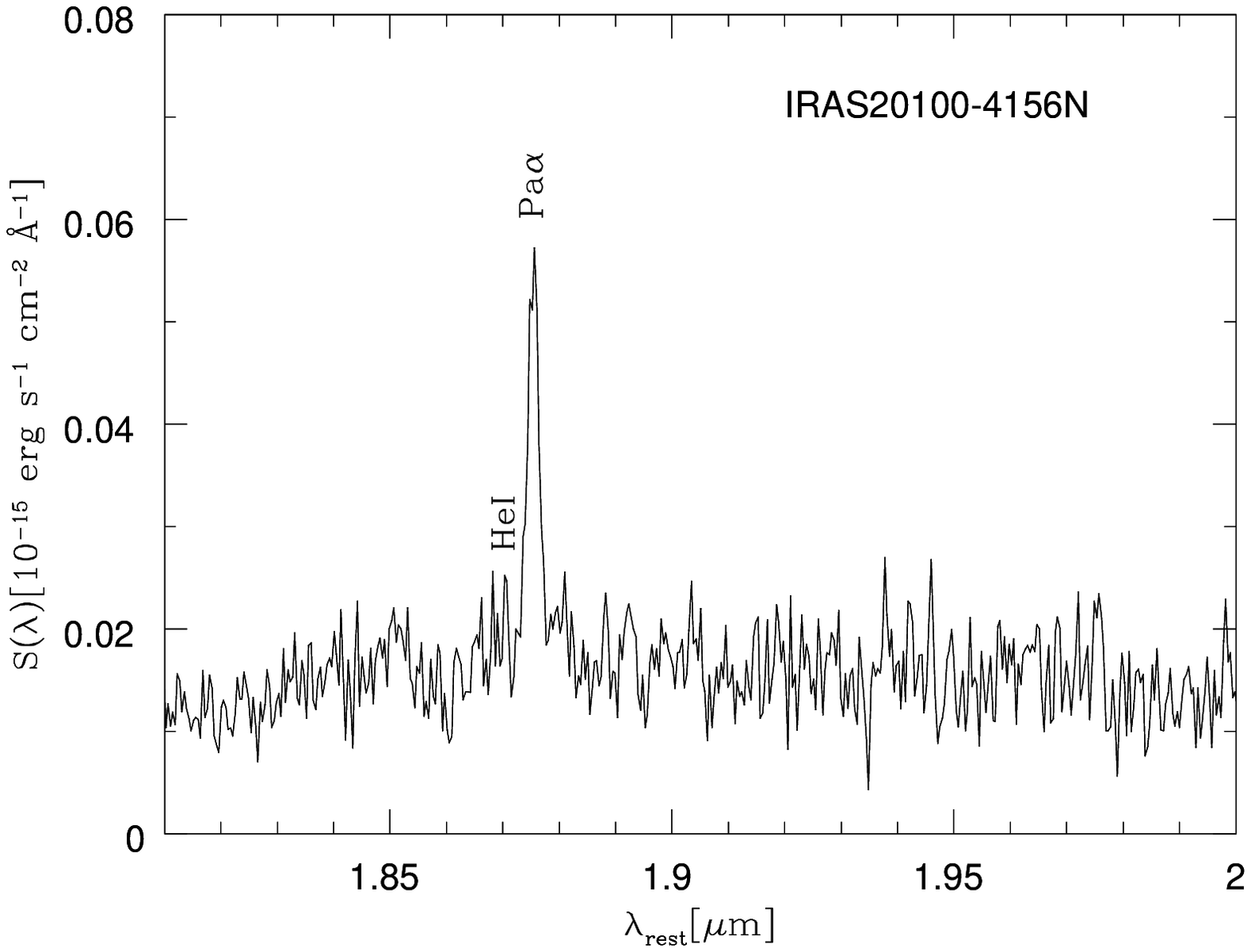}
                      \includegraphics{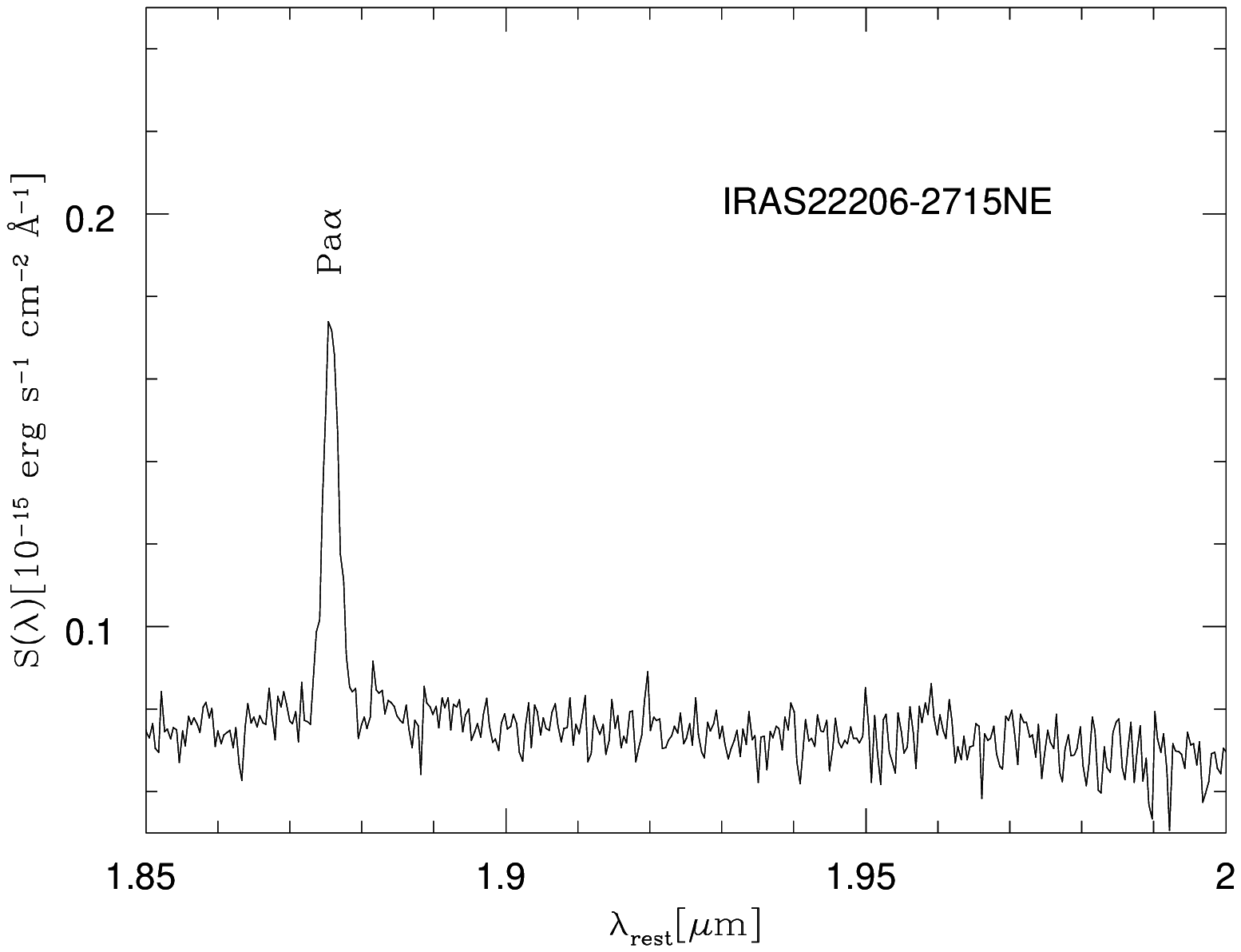}
                     \includegraphics{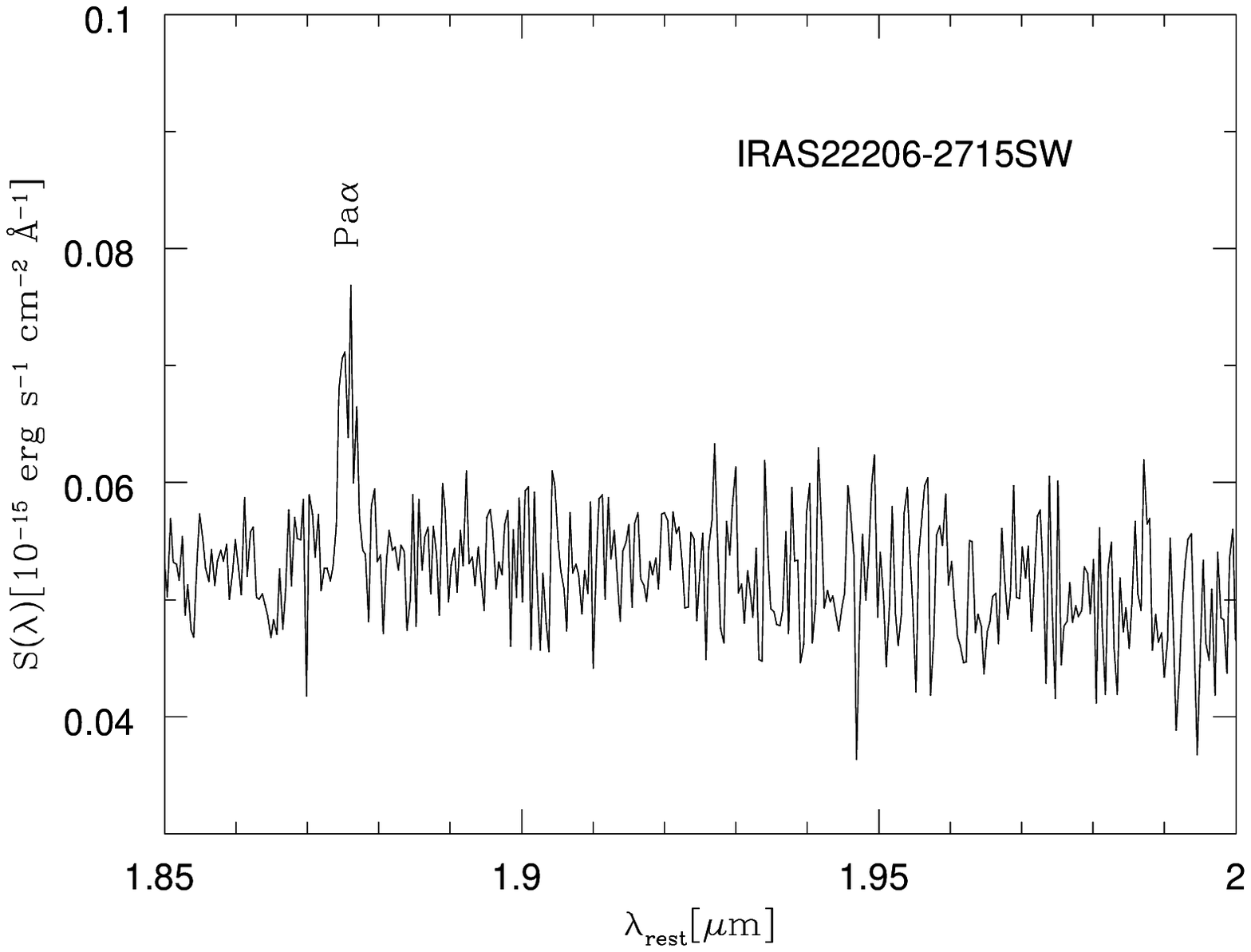}}
\caption{\footnotesize{Spectra of the sample galaxies
showing emission in two nuclei. No slit-losses, 
nor extinction correction have been applied. The identified emission
features are highlighted. Fluxes are in units of $[10^{-15}$erg s$^{-1}$cm
$^{-2}$\AA$^{-1}]$}.}
\label{spec2}
\end{figure*}

\begin{table*}[!ht]
\begin{flushleft}
\caption{Infrared and radio properties of the observed galaxies}
\label{irprop}
\begin{tabular}{lccccccccccc}\hline \hline
Object & z & D & S$_{12\mu m}$ & S$_{25\mu m}$ & S$_{60\mu m}$ &
 S$_{100\mu m}$ & S$_{25}$/S$_{60}$ & S$_{IR}$ & S$_{Radio}$  & q$^d$ & L/C~$^e$\\
& & [Mpc] & [Jy] & [Jy] & [Jy] & [Jy] & & [W~m$^2$] & [mJy] &  & 7.7$\mu$m\\ \hline
IRAS00085-1223 & 0.0198 & 80.3 & 0.395 & 2.372 & 16.62 & 16.97 & 0.143 &1.39$\times$10$^{-12}$ &  66.9$^b$  & 2.48& \\     
IRAS00188-0856 & 0.1284 & 648.6 & 0.117 & 0.372 & 2.59 & 3.40 & 0.144 & 2.44$\times$10$^{-13}$ &  15.7$^b$  & 2.36& 1.53 \\
IRAS00582-0258 & 0.0874 & 373.7 & 0.110 & 0.341 & 1.21 & 0.99 & 0.282 & 1.33$\times$10$^{-13}$ &  10.1$^b$  & 2.14& \\     
IRAS01077-1707 & 0.0334 & 158.4 & 0.301 & 0.846 & 6.48 & 10.4 & 0.130 & 6.40$\times$10$^{-13}$ &  43.8$^b$  & 2.32& \\     
IRAS02411+0354 & 0.1436 & 638.1 & 0.085 & 0.224 & 1.37 & 1.94 & 0.164 & 1.40$\times$10$^{-13}$ &  6.6$^b$   & 2.44& 3.17\\  
IRAS06206-6315 & 0.0924 & 455.9 & 0.069 & 0.294 & 3.96 & 4.58 & 0.074 & 3.10$\times$10$^{-13}$ & 21.9$^a$   & 2.53& 3.69 \\
IRAS19335-3632 & 0.0821 & 349.8 & 0.107 & 0.154 & 1.19 & 1.78 & 0.129 & 1.27$\times$10$^{-13}$ &  7.1$^b$   & 2.36&  \\    
IRAS20100-4156 & 0.1295 & 655.0 & 0.135 & 0.343 & 5.23 & 5.16 & 0.065 & 4.00$\times$10$^{-13}$ &  20.3$^c$  & 2.49& 1.92\\  
IRAS22206-2715 & 0.1314 & 579.1 & 0.096 & 0.160 & 1.75 & 2.33 & 0.091 & 1.61$\times$10$^{-13}$ &  6.3$^b$   & 2.56&  \\    
IRAS22491-1808 & 0.0777 & 379.7 & 0.119 & 0.549 & 5.44 & 4.45 & 0.101 & 4.13$\times$10$^{-13}$ &  5.9$^b$   & 3.02& 2.85 \\
IRAS23128-5919 & 0.0446 & 184.7 & 0.250 & 1.590 & 10.80 & 10.99 & 0.147 &9.07$\times$10$^{-13}$ & 37.3$^a$   & 2.72&  \\       
IRAS23230-6926 & 0.1063 & 529.2 & 0.058 & 0.295 & 3.74 & 3.42 & 0.079 & 2.77$\times$10$^{-13}$ & 32.0$^a$   & 2.31& 1.50 \\
IRAS23389-6139 & 0.0927 & 457.5 & 0.063 & 0.244 & 3.63 & 4.26 & 0.067 & 2.83$\times$10$^{-13}$ & 163.9$^a$  & 1.62& 1.33 \\
\hline
\end{tabular}                                                                                       
\footnotesize
a) 843MHz flux from Mauch et al (2003);
b) 1.4GHzflux from NRAO VLA Sky Survey (Condon et al 1998); c)
Condon et al (1996)\\
d) FIR/Radio correlation parameter (Sanders \& Mirabel 1996). 
843MHz fluxes were converted to 1.4GHz fluxes
by assuming F($\nu)\propto \nu^{-0.8}$\\
e) L/C from Lutz \et \cite{Lutz98}
\end{flushleft}
\end{table*}

\section{Observations and Data Reduction}

\label{obs}

The selected galaxies were drawn from the IRAS 1.2Jy sample (Fisher \et \cite{
fish}), and from Genzel \et (\cite{gen}) and Rigopoulou \et (\cite{rig1}). The
sample galaxies cover the interval of solar luminosities 11.45$\leq$log(L$_{IR}$/L
$_{\odot}$)$\leq$12.73.
The IR and radio properties of our target galaxies are listed in Table \ref{irprop}.
The penultimate  column indicates the FIR/Radio correlation parameter
at 1.4GHz, (see eq. \ref{qeq} below), which for star forming galaxies
is known to be q$\simeq$2.35$\pm$0.15 over a wide range of infrared luminosities
(e.g. Sanders \& Mirabel \cite{sand1}).

Medium-resolution Ks-band NIR spectroscopy of the galaxies  was performed with
the SOFI infrared spectrometer (Moorwood \et \cite{moor}) at the Nashmyth A focus
of the ESO 3.5m New Technology Telescope (NTT), during August 2001 and August
2003. During the 2003 run, 5 minutes of Ks-band imaging were obtained as
well.

The instrumental set consisted of the medium resolution grism (pixel scale =
0.292 arcsec), with a dispersion of 4.62 \AA/pixel and a spectral coverage between
2.0 $\mu$m and 2.3 $\mu$m. The Ks filter works as order sorting filter. We
adopted 1 arcsec slit, providing a spectral resolution of R~$\equiv$~$\Delta$
$\lambda$/$\lambda$~$\approx$~ 2000 ($\Delta$$v~\approx$~150 km~s$^{-1}$) at 2
$\mu$m.

Data were acquired by placing the galaxy at two different positions along the
slit, through the standard {\em nodding} technique; the nod throw was set to 60
arcsec.

In some cases, given the peculiar structure of the galaxy, the slit has been
oriented with two different position angles (see Table \ref{obs}).

The standard reduction pipeline for long-slit NIR spectra was performed by using
 routines of the Image Reduction and Analysis Facility (IRAF\footnote{The
 package IRAF is distributed by the National Optical Astronomy Observatory which
 is operated by the Association of Universities for Research in Astronomy, Inc.,
 under cooperative agreement with the National Science Foundation.}). The spectra
 were flat-field corrected in the usual manner and sky was subtracted through the
 {\em nod on slit} procedure.

Wavelength calibration and slit-curvature correction was obtained with the Xenon/
Neon internal lamp and checked on sky IR emission lines.

Corrections for telluric features and flux calibration have been performed by
observing nearly featureless hot B and/or solar-like telluric and spectro-
photometric standard stars, from the Hipparcos catalogue. Atmospheric
transmission variations were compensated by observing the standard stars
contiguously, at similar (as near as possible to target galaxies) air masses. The
adopted procedure consists in dividing the raw galaxies spectra by that of the
standard star and then multiplying  by its intrinsic emitted spectrum. The latter
has been obtained by normalizing Pickles (\cite{pick}) stellar templates of the
proper spectral type to the V-to-K observed magnitudes of the standard stars. In
the case of not optimal {\em seeing} conditions, we corrected the raw spectra of
the standard stars for slit-losses, before applying flux calibration.

The exposure time, the position angle (PA) of the slit, the spectral type of the
atmospheric calibrators and their air masses with respect to those of the
observed galaxies are given in Table \ref{obs}.

\begin{table}[!ht]
\begin{center}
\caption{Observational Parameters}
\label{obs}
\begin{tabular}{lrrcc}\hline \hline
IRAS name &  Integ. & P.A. & Atmos. & Gal./Std.\\
& (seg) & (deg) & Calib. & Air Mass\\ \hline
00085-1223 & 3600 & 38 & B9V & 1.11/1.11\\
00188-0856 & 7200 & 0 & F8V & 1.13/1.46\\
00582-0258 & 3600 & 55 & G0V & 1.18/1.17\\
               & 3600 & 67 & B9V & 1.11/1.16\\
01077-1707 & 480 & 30 & F6V & 1.22/1.25\\
02411+0354 & 6600 & -54 & G0V & 1.22/1.07\\
               & 6600 & 90 & G0V & 1.19/1.07\\
06206-6315 & 2400 & 145 & G0V & 1.57/1.69\\
19335-3632 & 5400 & 43 & B9V & 1.19/1.47\\
20100-4156 & 7500 & 25 & G2V & 1.32/1.05\\
               & 8400 & -56 & B9V & 1.11/1.12\\
22206-2715 & 3600 & -32 & G0V & 1.11/1.04\\
               & 3600 & 61 & G0V & 1.37/1.04\\
22491-1808 &  7200 & 74 & G3V & 1.02/1.08\\
23128-5919 & 4800 & 35 & B4V & 1.32/1.10\\
23230-6926 &  4500 & 158 & F7V & 1.45/1.51\\
23389-6139 &  480 & 8 & F5V & 1.20/1.03\\\hline
\end{tabular}
\end{center}
\end{table}

\subsection{Emission line properties}
\label{elprop}

NIR spectra of galaxies in our sample, presented in Figures
\ref{spec1} and \ref{spec2},  are characterized by strong
Pa$\alpha$ or Br$\gamma$ emission, depending on their redshift. 
In the Pa$\alpha$ region we also detect weak H$_2$ lines from higher order vibrational
transitions (H$_2$ 1-0S(5) at 1.835$\mu$m, H$_2$ 1-0S(3) at 1.957$\mu$m, and H$_2$
1-0S(2) at 2.033$\mu$m). Almost all these galaxies show a contamination of the Pa$\alpha$
blue wing by the HeI emission at 1.868$\mu$m, and in some of them the weak Br$\delta$
emission at 1.945$\mu$m is detected.
Galaxies observed in the Br$\gamma$ region, show H$_2$ lines from
lower order vibrational transitions (H$_2$ 1-0S(3) at 1.957$\mu$m, H
$_2$ 1-0S(2) at 2.033 $\mu$m, H$_2$ 1-0S(1) at 2.121$\mu$m, and H
$_2$ 1-0S(0) at 2.223$\mu$m). It is to note the relatively
strong HeI emission at 2.058$\mu$m in the case of IRAS 23128-5919.

\begin{figure}[!ht]
\label{lirfir}
\includegraphics[width=0.48\textwidth]{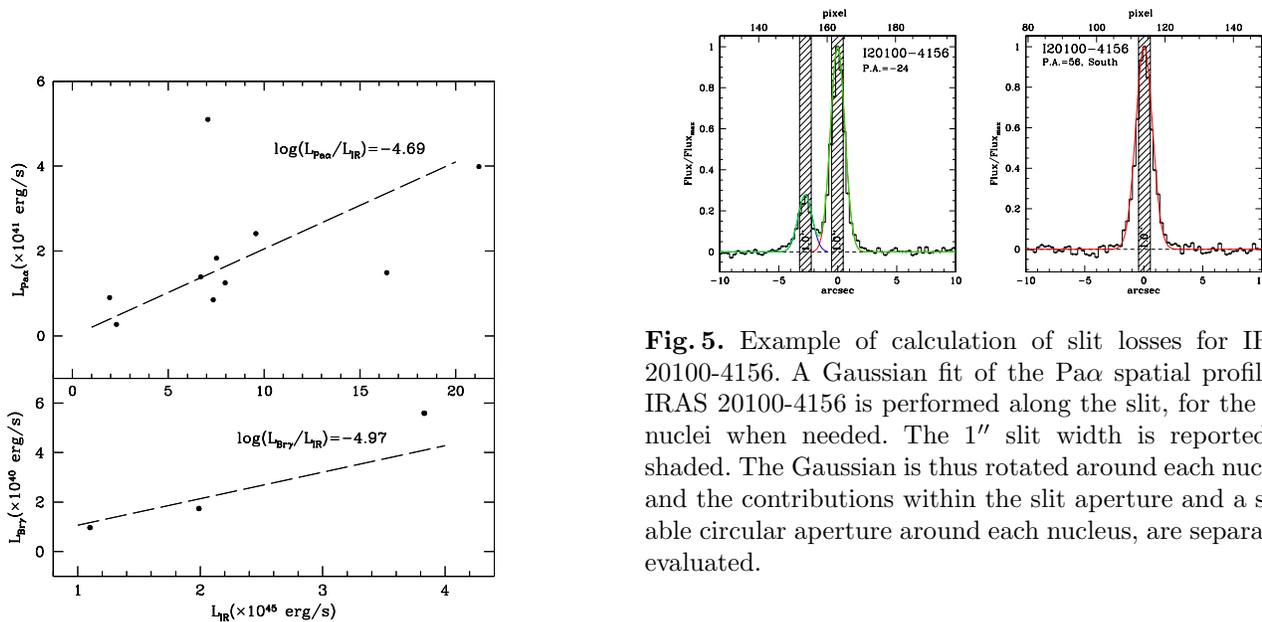}
\caption{Relation between \pa\ and IR luminosity (upper panel) and
\bg\ and IR luminosity (Lower panel). 
Dashed line in the upper panel is our linear relation
L(line)= Const $\times$ L(IR), while that in the lower panel
has ben taken from
the value quoted by Goldader et al. 1997.}
\label{filifir}
\end{figure}

We find a good correlation between the \pa\ and IR luminosity, as shown in Figure
\ref{filifir}. The dashed line in the Figure is
the zero-intercept linear correlation we obtain,
log(L$_{Pa\alpha}$/L$_{IR}$) = -4.69. 
IRAS 02411+0354 has been excluded from the analysis because
it falls significantly off the correlation.
In the case of  \bg\
luminosity, we have only three galaxies, 
but our values fall on top of the relation found
by Goldader \et (\cite{gold97}). 
The dashed line is a zero-intercept linear correlation with slope
quoted by Goldader \et (\cite{gold97}), log(L$_{Br\gamma}$/L$_{IR}$) = -4.96.

The spatial distribution of the \pa\ emission in
IRAS 00582-0258, IRAS 23230-6926
and IRAS 23389-6139 appears compact. The
peak  of the Gaussian profile of the line emission corresponds to the
peak obtained from the continuum emission alone,
coming from the nucleus of the galaxies. In contrast, IRAS 22491-1808 shows a
broader spatial distribution in  Pa$\alpha$ emission, with some minor peaks
superposed on a broader single Gaussian profile along the slit.

Six galaxies, IRAS 00188-0856, IRAS 02411+0354, IRAS 19335-3632, IRAS 20100-4156, IRAS
22206-2715, and IRAS 22491-1808 show double profiles, corresponding to the
double nuclei observed in the corresponding NIR images. 
Finally,  IRAS 06206-6315 is best fitted 
with three Gaussian profiles. This complexity is associated with the
bright regions seen in the H and I images (Bushouse \et \cite{bush}).

In the case of  galaxies observed in the Br$\gamma$ spectral domain,
IRAS00085-1223 and IRAS23128-5919 show a single Gaussian profile,
while  IRAS 01077-1707 is best fitted 
with three Gaussian profiles. 

The observed emission line properties of target galaxies are shown in Table
$\ref{elines}$. Only for those galaxies with sufficient S/N have we considered the
emission properties of multiple components separately while, in all other cases, we have
folded the separate contribution together. The FWHM velocities have been corrected
by instrumental response after subtracting in quadrature the instrumental FWHM.

\subsection{Detection of AGN signatures}
\label{agnsig}

Typical signatures  of central non thermal activity are the presence
of broad emission line components and/or high excitation lines.

None of the objects observed in the \bg\ region show a broad
component of the \bg\ line. 
In the case of the \pa\ galaxies, we have attempted to isolate a broad component of the
Pa$\alpha$ emission, by forcing the fit of the shape of the line with two Gaussians. With
this procedure, in IRAS 00188-0856 and IRAS 00582-0258 we could detect a broad component
of FWHM $\simeq$ 2339 and  2210 km/s, respectively. Figure \ref{broad} shows the
results of the multi Gaussian fitting for these galaxies. In the case of IRAS 00188-0856
we have also taken into account the contribution of the HeI line at 1.8689$\mu$m to the
general shape of the line. The broad component corresponds to about 42\% and 80\% of the
Pa$\alpha$ flux in the narrow component for IRAS 00188-0856 and IRAS 00582-0258,
respectively. The Pa$\alpha$ intensities reported in Table \ref{elines} correspond only
to the narrow component.

\begin{figure*}[!ht]
\centering
\label{broad}
\includegraphics[width=0.48\textwidth]{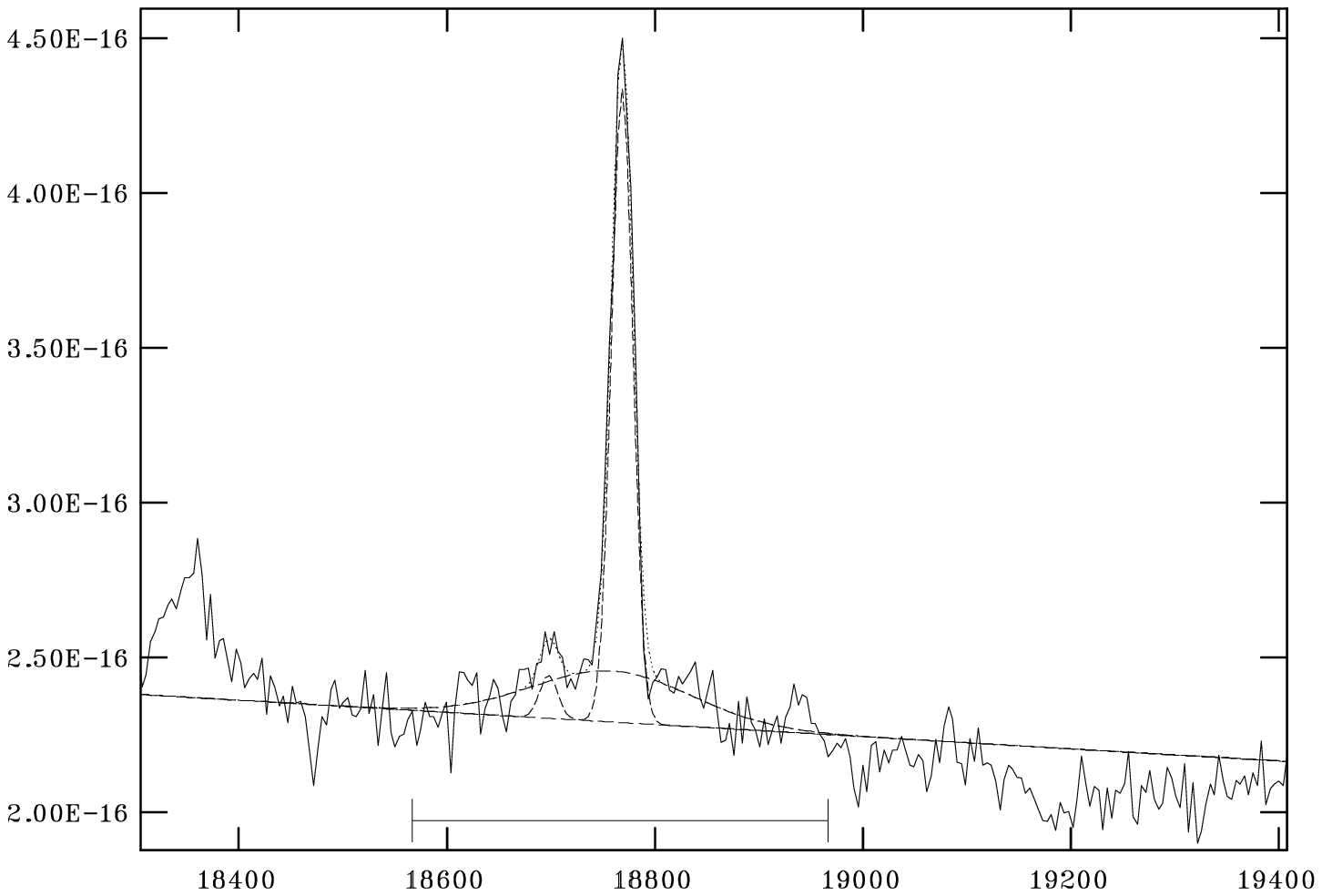}
\includegraphics[width=0.48\textwidth]{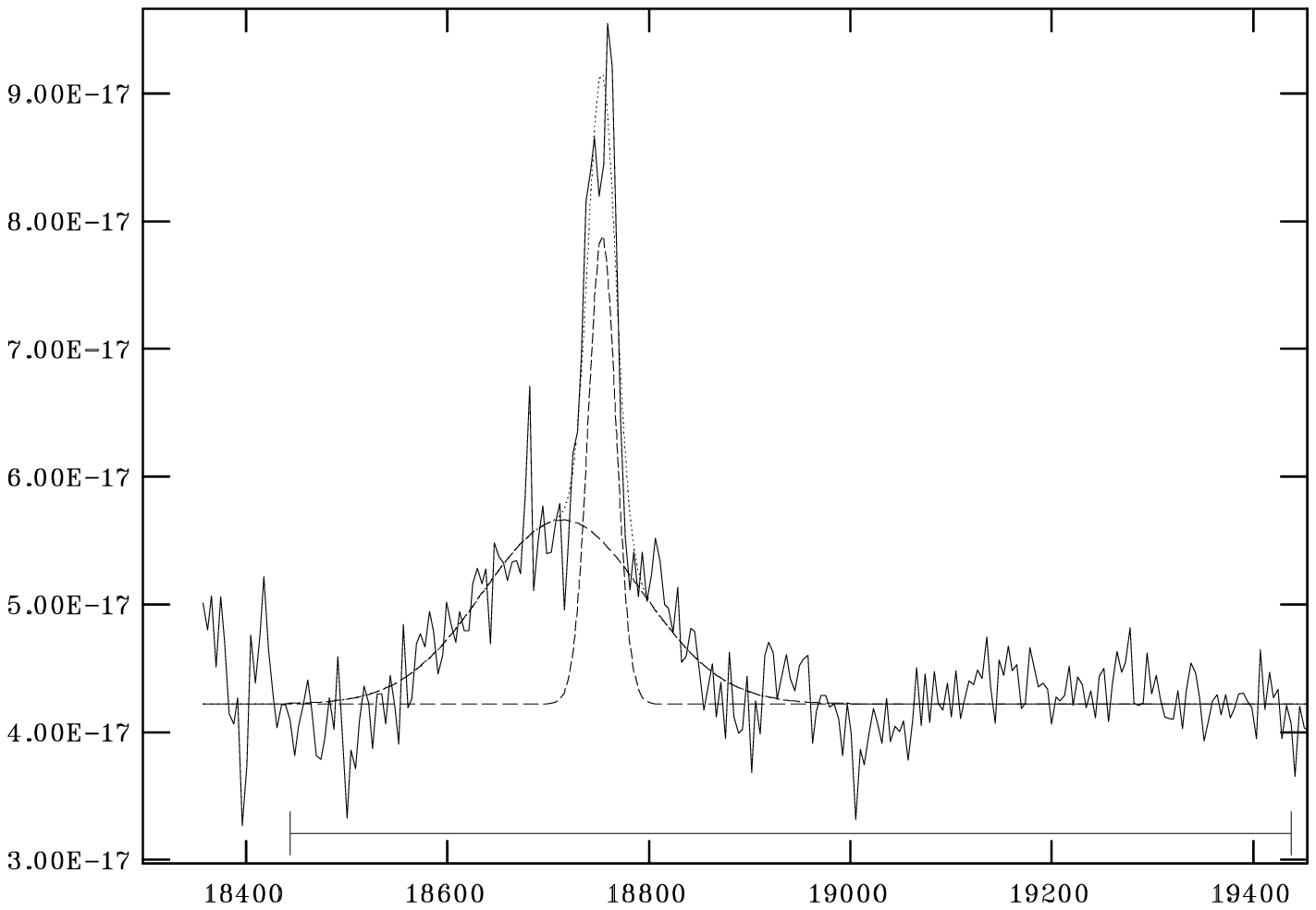}
\caption{Possible detection of a broad emission component in the IRAS 00188-0856 (left panel) and IRAS 00582-0258 (right panel). The width of the broad component is
FWHM$\simeq$2339 km/s for IRAS 00188-0856, and
FWHM$\simeq$2210 km/s for IRAS 00582-0258.}
\label{broad}
\end{figure*}

Another AGN diagnostic in the NIR, is the presence of the high
excitation line [SiVI]$\lambda$1.9628$\mu$m.  
This diagnostic is particularly useful in Sy2 galaxies,
that lack broad-line components. The [SiVI]$\lambda$1.9628$\mu$m line falls slightly
long wards of the H$_2$ 1-0S(3) 1.957 $\mu$m
line, causing an asymmetry in medium resolution spectra (e.g. Vanzi \et \cite{
vanzi}). While our spectral resolution allow us to disentangle these two lines,
the S/N is not sufficient to draw firm conclusions. However, [SiVI]$\lambda$1.9628$\mu$m
seems to be present in the spectra of IRAS 00188-0856  and IRAS 00582-0258,
the same galaxies for which we have hints of a broad emission component
at Pa$\alpha$. In all other cases there is no evidence of this coronal line.

\begin{table*}[!ht]
\begin{center}
\caption{Observed Emission Lines Properties of Target Galaxies}
\label{elines}
\begin{tabular}{lcccrrr}\hline \hline
IRAS name & Line & $\lambda_0$ & Flux & EW & FWHM & FWHM \\
& & [$\mu$m] & [erg s$^{-1}$ cm$^{-2}$] & [rest$\AA$] & [rest$\AA$] & [km$^{-1}$]\\ \hline

00085-1223 & HeI & 2.1500 & 4.99$\times$10$^{-15}$ & -1.04 & 34.26 & 482.26\\
& Br$\gamma$ & 2.1661 & 1.25$\times$10$^{-14}$ & -2.68 & 35.51 & 475.85\\
& HeII & 2.2155 & 4.34$\times$10$^{-15}$ & -0.99 & 29.97 & 348.24\\
& H$_2$ 1-0S(0) & 2.2235 & 6.15$\times$10$^{-15}$ & -1.43 & 20.45 & 243.83\\ \hline

00188-0856 & H$_2$ 1-0S(5)& 1.8353 & 4.21$\times$10$^{-16}$ & -1.81 & 16.86 & 243.82\\
& HeI & 1.8689 & 3.86$\times$10$^{-16}$ & -1.79 & 18.87 & 278.64\\
& Pa$\alpha$ & 1.8756 & 5.97$\times$10$^{-15}$ & -29.4 & 27.88 & 423.43\\
& H$_2$ 1-0S(3) & 1.9570 & 5.01$\times$10$^{-16}$ & -2.63 & 16.45 & 215.47 \\ \hline

00582-0258 & HeI & 1.8689 & 9.34$\times$10$^{-17}$ & -1.89 & 4.37 &\\
& Pa$\alpha$ & 1.8756 & 1.63$\times$10$^{-15}$ & -31.56 & 36.19 & 561.35\\
& H$_2$ 1-0S(3) & 1.9570 & 2.05$\times$10$^{-16}$ & -3.18 & 24.26 & 294.90\\
\hline

01077-1707 & H$_2$ 1-0S(2) & 2.0338 & 1.29$\times$10$^{-15}$ & -1.89 & 18.27 & 218.47\\
& HeI & 2.0587 & 2.22$\times$10$^{-15}$ & -3.27 & 26.72 & 409.34\\
& H$_2$ 1-0S(1) & 2.1213 & 2.16$\times$10$^{-15}$ & -3.45 & 23.68 & 290.23\\
& Br$\gamma$ & 2.1661& 5.61$\times$10$^{-15}$ & -9.38 & 33.43 & 514.02\\ \hline

02411+0354NE & HeI & 1.8150 & 3.32$\times$10$^{-15}$ & -6.95 & 23.00 & 347.08\\
& H$_2$ 1-0S(5) & 1.8353 & 1.56$\times$10$^{-16}$ & -2.95 & 17.62 & 251.22\\
& HeI & 1.8689 & 5.25$\times$10$^{-16}$ & -8.58 & 26.38 & 420.41\\
& Pa$\alpha$ & 1.8756 & 8.30$\times$10$^{-15}$ & -151.5 & 22.67 & 339.52\\
& H2 1-0S(4) & 1.8920 & 2.52$\times$10$^{-16}$ & -4.82 & 19.11 & 305.08\\
& Br$\delta$ & 1.9445 & 3.29$\times$10$^{-16}$ & -6.03 & 17.39 & 272.50\\
& H$_2$ 1-0S(3) & 1.9570 & 3.73$\times$10$^{-16}$ & -5.59 & 21.31 & 301.19\\ \hline

02411+0354SW & Pa$\alpha$ & 1.8756 & 2.18$\times$10$^{-15}$ &-92.3 & 24.37 & 369.74\\ \hline

06206-6315 & Pa$\alpha$ & 1.8756 & 5.05$\times$10$^{-15}$ & -30.25 & 42.01 & 651.92\\
& H$_2$ 2-1S(6) & 1.8942 & 5.66$\times$10$^{-16}$ & -3.60 & 11.15 & 137.46\\ \hline

19335-3632 & Pa$\alpha$ & 1.8756 & 6.15$\times$10$^{-15}$ & -124.0 & 24.64 & 368.26 \\
& Br$\delta$ & 1.9445 & 5.19$\times$10$^{-16}$ & -7.43 & 17.93 & 228.89\\
& H$_2$ 2-1S(3) & 1.9570 & 4.07$\times$10$^{-16}$ & -5.82 & 18.79 & 264.55\\
& HeI & 2.0587 & 4.71$\times$10$^{-15}$ & -8.78 & 18.23 & 237.63\\ \hline

20100-4156S & H$_2$ 2-1S(5) & 1.8353 & 3.1$\times$10$^{-16}$ & -11.63 & 30.25 & 497.00\\
& HeI & 1.8689 & 8.8$\times$10$^{-16}$ & -18.61 & 62.87 & 1008.76\\
& Pa$\alpha$ & 1.8756 & 6.09$\times$10$^{-15}$ & -136.2 & 31.36 & 501.56 \\
&Br$\delta$ & 1.9445 & 6.00$\times$10$^{-16}$ & -13.91 & 35.02 & 540.18\\
& H$_2$ 2-1S(3) & 1.9570 & 1.66$\times$10$^{-15}$ & -39.04 & 51.47 & 788.84\\ \hline

20100-4156N & HeI & 1.8689 & 3.29$\times$10$^{-16}$ & -14.08 & 33.40 & 535.97\\
& Pa$\alpha$ & 1.8756 & 1.70$\times$10$^{-15}$ & -76.95 & 25.46 & 407.21 \\ \hline

22206-2715NE & Pa$\alpha$ & 1.8756 & 2.81$\times$10$^{-15}$ & -37.78 & 26.67 & 408.40\\ \hline

22206-2715SW & Pa$\alpha$ & 1.8756 & 6.58$\times$10$^{-16}$ & -13.18 & 28.79 & 439.67\\ \hline

22491-1808 & Pa$\alpha$ & 1.8756 & 4.94$\times$10$^{-15}$ & -90.96 & 30.07 & 449.54\\
& Br$\delta$ & 1.9445 & 1.21$\times$10$^{-16}$ & -2.00 & 12.26 & 142.29\\
& H$_2$ 1-0S(3) & 1.9570 & 7.04$\times$10$^{-16}$ & -11.93 & 34.18 & 513.09\\
& H$_2$ 1-0S(1) & 2.1213 & 8.31$\times$10$^{-16}$ &-18.75 & 33.77 & 459.81\\\hline

23128-5919N & HeI & 2.0587 & 3.75$\times$10$^{-15}$ & -8.30 & 25.75 & 427.81\\
& H$_2$ 1-0S(2) & 2.0338 & 8.28$\times$10$^{-15}$ & -18.48 & 24.44 & 328.09\\
& H$_2$ 1-0S(1) & 2.1213 & 5.85$\times$10$^{-15}$ & -13.99 & 27.87 & 375.23\\
& Br$\gamma$ & 2.1661 & 1.37$\times$10$^{-14}$ & -30.78 & 57.11 & 811.43\\ \hline

23230-6926 & Pa$\alpha$ & 1.8756 & 7.21$\times$10$^{-15}$ & -111.8 & 35.53 & 501.97\\
& Br$\delta$ & 1.9445 & 5.93$\times$10$^{-16}$ & -8.94 & 20.70 & 303.25\\
& H$_2$ 1-0S(3) & 1.9570 & 1.30$\times$10$^{-15}$ & -21.94 & 33.92 & 497.54\\
& H$_2$ 1-0S(2) & 2.0338 & 2.87$\times$10$^{-16}$ & -4.45 & 15.13 & 174.93\\ \hline

23389-6139 & Pa$\alpha$ & 1.8756 & 7.34$\times$10$^{-15}$ & -77.48 & 49.56 & 676.14\\
& H$_2$ 2-1S(6) & 1.8942 & 1.71$\times$10$^{-16}$ & -1.72 & 4.65 & \\
& Br$\delta$ & 1.9445 & 9.30$\times$10$^{-17}$ & -0.90 & 2.76 & \\
& H$_2$ 1-0S(3) & 1.9570 & 2.38$\times$10$^{-15}$ & -27.34 & 43.44 & 639.7\\
& [FeII] & 1.9670 & 2.63$\times$10$^{-16}$ & -2.89 & 15.24 & 150.4\\ \hline
\end{tabular}
\end{center}
\end{table*}

\section{The star formation rates.}
\label{sfrates}

In this section we compare the star formation rate derived from the \pa\ or \bg\
emission lines, with that deduced from the IR, assuming that
both  originate in an ongoing vigorous starburst.

In order to perform a proper analysis of the star formation rate
from the emission lines we need to correct the measured
fluxes for slit losses and for dust extinction.

\subsection{Slit loss corrections}
\label{los}

When comparing different observations, it is important to
evaluate any discrepancy caused by the different apertures. If the source
has a certain degree of symmetry, it is possible to extrapolate the mono
dimensional spatial profile observed along the slit, to a two-dimensional surface
brightness distribution, and to estimate the flux that
would be recevied by a given aperture.
We define as ``slit losses'' the ratio between the
flux corresponding to a given aperture and the flux received within our
rectangular aperture. 

\begin{figure}[!h]
\centering
\includegraphics[width=0.22\textwidth]{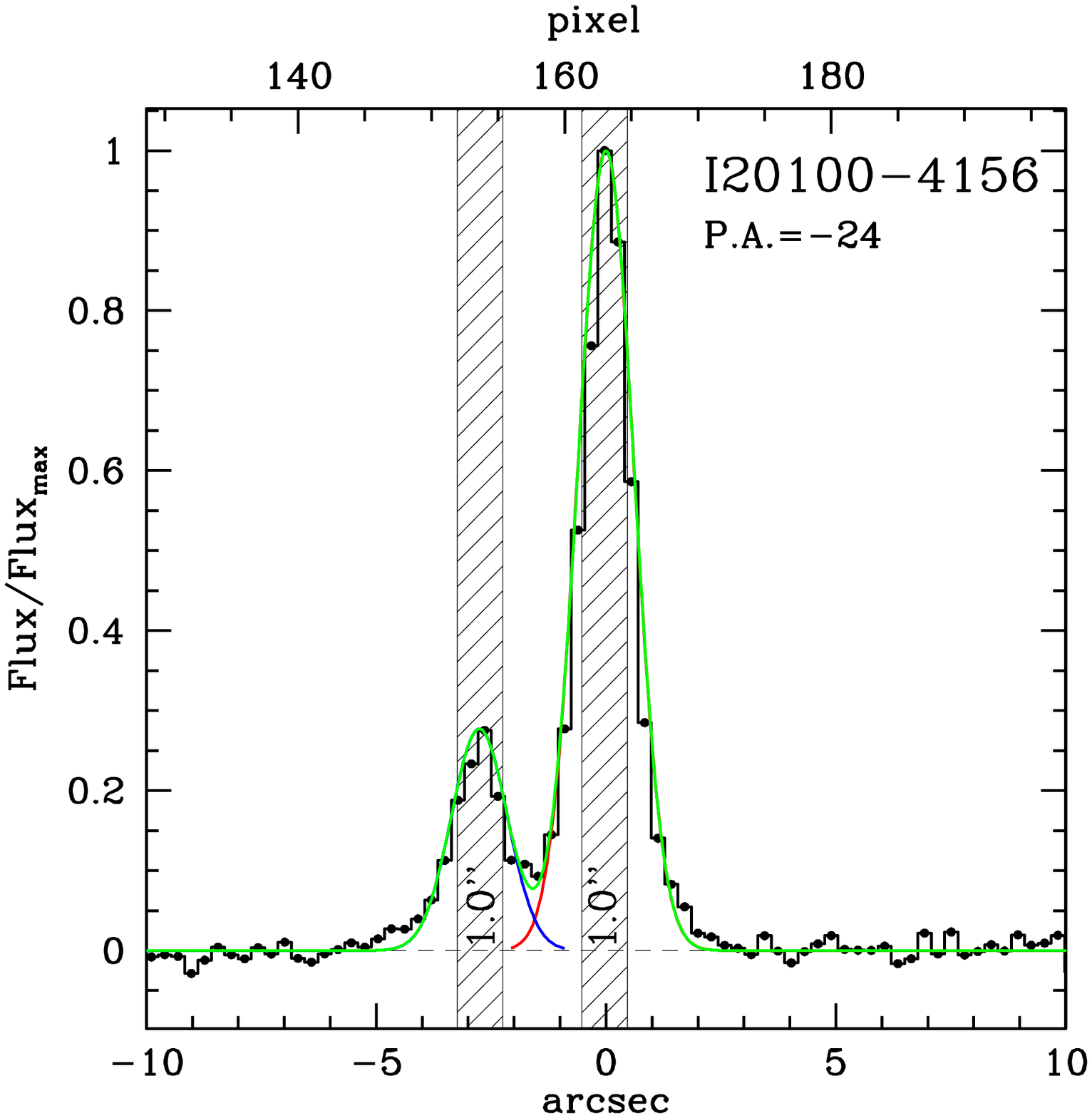}
\includegraphics[width=0.22\textwidth]{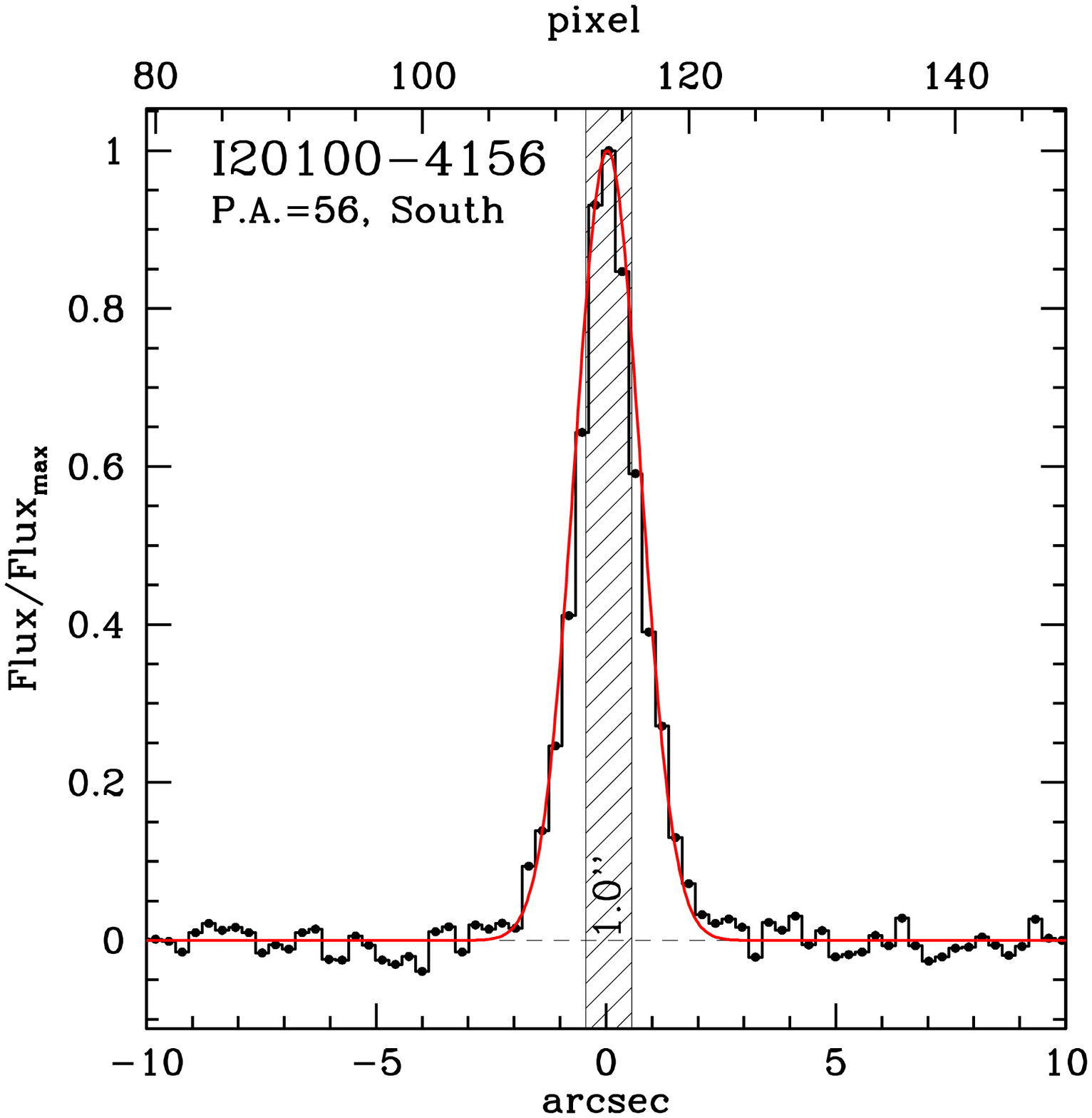}
\caption{Example of calculation of slit losses for IRAS 20100-4156.
A Gaussian fit of the
\pa\ spatial profile of IRAS 20100-4156 is performed along the slit, for the two
nuclei when needed. The 1$''$ slit width is reported as shaded. The Gaussian is
thus rotated around each nucleus and the contributions within the slit aperture
and a suitable circular aperture around each nucleus, are separately evaluated.}
\label{spprof}
\end{figure}

In our  galaxies,  the two-dimensional flux distribution was obtained by rotating
around the nucleus the best fit Gaussian profile of the spatial distribution of
the \pa\ or \bg\ line.
If the slit covers more than one nucleus, different
Gaussian profiles were fitted for each nucleus. As an example, Figure \ref{spprof} 
shows the case of IRAS20100-4156, at two position angles
P.A.=-24 and 56 respectively.
Slit loss coefficients,$C_{H\alpha}$ and $C_{IR}$, 
needed to compare our \pa\ fluxes to \ha\ and IR fluxes
taken from the literature are 
reported in Table \ref{losses}.

\subsection{Attenuation}
\label{att}

Correction for attenuation can be derived by comparing the observed hydrogen
line emission intensity ratios, R$_{obs}$, with the intrinsic values predicted
by models of nebular emission, R$_{int}$.

However, since the only strong Hydrogen line present in our spectra
is either \pa\ or \bg\, we need to compare our fluxes with other observations 
taken from the literature.
Duc, Mirabel \& Maza (\cite{duc}), Veilleux, Kim \& Sanders (\cite{veil99}), and
Veilleux \et (\cite{Veil95}) provide optical spectroscopy within slit apertures of
1.25$''$, 2 kpc and 4 kpc respectively, for almost all galaxies in our sample.
Observed intensities of the \ha\ and \hb\ emission lines
are reported in columns two and three of Table \ref{ext}.

There exists a wide literature concerning predicted intrinsic line ratios of Hydrogen
recombination lines. In general these values depend only very slightly on the assumed
density of the emitting region (usually between $10^2-10^4$cm$^{-3}$) but they depend
more on the electronic temperature. The latter may be assumed or may be the result of
detailed modeling of HII regions, accounting for the hardness of ionizing spectrum,
geometry and metallicity of the gas. To investigate this point, we have run a set of
CLOUDY (Ferland, 2003) models for different HII regions, characterized by different ages of the
ionizing stellar population, different metallicity and gas density. The intrinsic
ratio decreases at increasing metallicity (lower electronic temperature) and
increasing age (lower hardness, neglecting a hardening of the spectrum due to the
presence of Wolf Rayet stars). For example, for the ratio  R
$_{int}^{H\alpha,Pa\alpha}$ we find values as high as 8.5 at Z=0.008
and young ages and as low as 7.0 in  metal rich HII regions  and/or old cluster ages.
Furthermore, more realistic galaxy models must take into
account a law of star formation 
(almost continuous in the case of a normal star forming
galaxy) which, in general will diminish the hardness of the ionizing spectrum. 
In all the following analysis we will make use of the recent
models for normal star forming galaxies presented by Panuzzo \et (\cite{pan03}). In
these models the galaxy evolution is followed with a chemical evolution code and the
spectro-photometric properties are computed with metal dependent stellar population
synthesis (Bressan, Chiosi \& Fagotto \cite{bress94}). Line emission is also accounted
for by means of a large data set of HII region models computed with CLOUDY, with the
appropriate chemical composition. The code (GRASIL, Silva \et \cite{silva98}) 
also account for attenuation of
light by dust, both in molecular clouds and in the more diffuse cirrus component.
These models have been successfully tested against a number of different observations,
from UV to the radio (Granato \et \cite{gran00}, Bressan \et
\cite{bress02} and Panuzzo \et \cite{pan03}).

We adopt the values quoted by Panuzzo \et (\cite{pan03}) for a star-forming
galaxy, i.e. R$_{int}^{H\alpha,H\beta}$$\simeq$3.0, R$_{int}^{H\alpha,Pa\alpha}$
$\simeq$7.2 and R $_{int}^{H\alpha,Br\gamma}$$\simeq$93.5. The latter two values,
in particular, are lower than that quoted by Hummer \& Storey (\cite{humm87}) (
$\simeq$8.58, and $\simeq$103.54), which however refer to a density of 10$^4$cm
$^{-3}$ and a temperature of Te=10$^4$K.  Our single population CLOUDY models
show that, already at Z=0.02 (solar), the temperature is below 8000K and e.g. R
$_{int}^{H\alpha,Pa\alpha}$$\simeq$8.0. Adopting R$_{int}^{H\alpha,Pa\alpha}$ by
Hummer \& Storey (\cite{humm87}), would increase the derived
E(B-V) by about 0.09 mag.

To evaluate the extinction from the observed \pa\ or \bg\ fluxes, we have re-scaled
our observations to the \ha\ aperture by multiplying our values by the coefficients
C$_{H\alpha}$ listed in Table \ref{losses}. As for the extinction law we have
used that of Calzetti et al. (2000), which is suited for starburst galaxies. The
derived
values of A$_V$ are shown in Table \ref{ext}.

\begin{table}[!ht]
\begin{center}
\caption{Coefficients for slit corrections.}
\label{losses}
\begin{tabular}{lcc}\hline \hline
IRAS name & C$_{H\alpha}$ & C$_{IR}$ \\\hline
00085-1223  & 1.67 & 2.08\\
00582-0258  & & 1.74\\
00188-0856  & 1.18 & 1.44\\
01077-1707  & 1.28 & 3.44\\
02411+0354NE  & 1.05 & 1.90\\
02411+0354SW  & 1.05 & 1.88\\
06206-6315  & 1.32 & 2.44\\
19335-3632  & & 3.56\\
20100-4156S  & 1.21 & 1.70\\
20100-4156N  & & 3.35\\
22206-2715  & 0.41 & 3.03\\
22206-2715NE  &  & \\
22206-2715SW  &  & \\
22491-1808  & 0.83 & 4.79\\
23128-5919  & 1.24 & 2.04\\
23230-6926  & 1.16 & 3.02\\
23389-6139  & 1.22 & 1.76\\
\hline
\end{tabular}
\end{center}
\end{table}

\subsection{The star formation rate.}

To derive the star formation rate from the observed line intensities
we have adopted the calibrations provided by Panuzzo \et (\cite{pan03}):

SFR(Pa$\alpha$) = 5.06$\times$10$^{-41}$~L(Pa$\alpha$)~M$_{\odot}$yr$^{-1}$/(erg\ s$^{-1}$)

SFR(Br$\gamma$) = 6.60$\times$10$^{-40}$~L(Br$\gamma$)~M$_\odot$yr$^{-1}$/(erg\ s$^{-1}$)

For the IR luminosity  (Table \ref{irprop})
we have adopted the calibration provided by the same authors
between the star formation rate and the 8\mum-1000\mum\ infrared luminosity:

SFR(IR) = 4.63$\times$10$^{-44}$~L(IR)~M$_\odot$yr$^{-1}$/(erg\ s$^{-1}$)

All the above calibrations refer to a Salpeter IMF between 0.1M$_{\odot}$ and 120M
$_{\odot}$ and take into account an increment of 16\% due to the lower IMF limit with
respect to the value of 0.15M$_{\odot}$ adopted by Panuzzo \et (\cite{pan03}). 
\pa\ and \bg\ fluxes were corrected for
extinction and multiplied by the coefficients 
C$_{IR}$ listed in Table \ref{losses}.
Luminosities have
been computed adopting  distances quoted in Table \ref{irprop}.

The values of SFR, obtained in this way from different indicators, are reported in Table
\ref{sfr}.
The last column of Table \ref{sfr} indicates the ratio between the
SFR derived from the hydrogen line emission and that derived from the
IR luminosity.

On average the \pa\ luminosity, even corrected for 
slit losses and attenuation, provides a SFR which is only 14\% of
that derived from the far infrared luminosity (assuming that the latter is 
entirely due to the starburst). In the case 
of the objects showing the \bg\ line, the average ratio is 
60\%.

\begin{table*}[!ht]
\begin{center}
\caption{Dust Extinction}
\label{ext}
\begin{tabular}{lccccc}\hline \hline
IRAS name & S$_{H\beta}$ & S$_{H\alpha}$ & \multicolumn{3}{c}{A$_V$}\\ \cline{4-6}
& & & H$\alpha$/H$\beta$ & Pa$\alpha$/H$\alpha$ & Br$\gamma$/H$\beta$\\ \hline
00085-1223 & 4.08e-15$^a$ &  1.00e-13$^a$ &  6.08 & &  5.09\\ 
00188-0856 & 2.45e-16$^b$ &  3.50e-15$^b$ &  4.52 & 4.28 & \\
01077-1707 & 1.16e-14$^a$ &  9.90e-14$^a$ &  3.03 & &  2.97\\ 
02411+0354NE & 8.00e-16$^b$ &  4.20e-15$^b$ &  1.62 & 4.33 & \\
02411+0354SW & 1.10e-15$^b$ &  7.60e-15$^b$ &  2.41 & 1.24 & \\
06206-6315 & 2.60e-16$^c$ &  6.31e-15$^c$ &  6.05 & 3.25 &\\
20100-4156S  & 1.86e-15$^c$ &  1.61e-14$^c$ &  3.06 & 1.91 & \\
22206-2715 & 8.96e-16$^b$ &  6.40e-15$^b$ &  2.51 & 0.75 & \\
22491-1808 & 3.15e-15$^b$ &  2.30e-14$^b$ &  2.57 & 0.40 & \\
23128-5919 & 8.86e-15$^c$ &  8.29e-14$^c$ &  3.29 & & 4.11\\
23230-6926 & 1.08e-15$^c$ &  1.35e-14$^c$ &  4.13 & 2.39 & \\
23389-6139 & 3.40e-16$^c$ &  2.26e-14$^c$ &  8.96 & 1.68 & \\\hline
\end{tabular}
\end{center}

\vspace{0.3cm}
\footnotesize
a) Veilleux \et (\cite{Veil95}); b) Veilleux, Kim \& Sanders (\cite{veil99}); 
c) Duc, Mirabel \& Maza (\cite{duc}) 

\end{table*}

\begin{table*}[!ht]
\begin{center}
\caption{SFR  derived from the observed
 \pa\ (or \bg) and IR luminosity. The spectral lines data have been
corrected for attenuation and for slit losses.}
\label{sfr}
\begin{tabular}{lrrrrrcrr}\hline \hline
IRAS name & L(Pa$\alpha$) & SFR(Pa$\alpha$) & L(Br$\gamma$) & SFR(Br$\gamma$) & L(IR) & log & SFR(IR)&  Line/IR \\
& [10$^{41}$erg s$^{-1}$] & [M$_{\odot}~$yr$^{-1}$] & [10$^{41}$erg s$^{-1}$] & [M$_{\odot}~$yr$^{-1}$] & [10$^{45}$erg s$^{-1}$] & (L$_{IR}$/L$_{\odot})$ & [M$_{\odot}~$yr$^{-1}$]\\ \hline
00085-1223 & & & 4.05 & 26.7 & 1.08 & 11.45 & 49.9 & 0.54   \\
00188-0856 & 9.1 & 48.1 & & & 12.32 & 12.51 & 570.5 & 0.08   \\
00582-0258 & 0.47 & 2.4 & & & 2.22 & 11.76 & 102.7 & 0.02    \\
01077-1707 & & & 8.22 & 54.2 & 1.92 & 11.70 & 89.1 & 0.61 \\
02411+0354 & 13.22 & 66.9 & & & 6.83 & 12.25 & 316.2 & 0.21   \\
02411+0354NE & 10.17 & 51.5 & & & &           &     \\
02411+0354SW & 3.04 & 15.4 & & & &            &       \\
06206-6315 & 8.78 & 44.4 & & & 7.73 & 12.30 & 357.9 & 0.12   \\
19335-3632 & 3.20 & 16.2 & & & 1.87 & 11.69 & 86.6    &  0.19  \\
20100-4156 & 12.00 & 60.7 & & & 20.59 & 12.73 & 953.1 &  0.06   \\
20100-4156S & 9.08 & 45.9 & & & &             &         \\
20100-4156N & 2.92 & 14.8 & & & &             &        \\
22206-2715 & 6.54 & 33.1 & & & 6.49 & 12.23 & 300.4   &  0.11 \\
22491-1808 & 6.38 & 32.3 & & & 7.13 & 12.27 & 330.1   &  0.10  \\
23128-5919 & & & 16.68 & 110.14 & 3.71 & 11.99 & 171.8   &  0.64 \\
23230-6926 & 15.01 & 75.9 & & & 9.29 & 12.38 & 430.0  &  0.18\\
23389-6139 & 15.42 & 78.0 & & & 7.11 & 12.27 & 329.0 & 0.24 \\ \hline
\end{tabular}
\end{center}

\vspace{0.3cm}
Pa$\alpha$ luminosities for IRAS00582-0258, IRAS19335-3632 and IRAS20100-4156N are not
extinction corrected, and the SFRs are only lower limits.
\end{table*}

\section{Discussion}
\label{disc}

In all the observed galaxies the \pa\ or \bg\ flux,
even corrected for aperture effects and extinction 
derived from NIR optical recombination lines  
(Table \ref{ext}), 
is significantly less
than that expected from a
starburst of corresponding bolometric luminosity.
A similar "deficit" of recombination photons has already been  
noticed by Goldader et al. (1995), based on the analysis of the Br$\gamma$ line
in a sample of local ULIRGs. 

Poggianti, Bressan \& Franceschini
(\cite{pogg}) have found that the SFR derived from the {\it extinction
corrected} intensity of \ha\ in a sample of ULIRGs was a factor of three
less than that derived from the FIR luminosity. They attributed the
discrepancy to an age-selective extinction effect where a substantial young
population was essentially escaping the optical detection. On the contrary,
in a recent analysis of lower infrared luminosity starburst galaxies (10
$\leq$log(L$_{IR})\leq$11) Mayya et al.(\cite{mayya}) have not found strong
evidence for a loss of optical photons.

We discuss below several possible explanations of the differences
between the SFR (and/or attenuation) derived from the emission lines and
from the IR emission.

\subsection{The AGN contribution}
\label{agn}

The first and most obvious cause for the photon deficit is that
a significant contribution to the infrared emission
comes from an AGN. There is evidence that in the most luminous nearby infrared
galaxies a significant fraction of the luminosity could arise in principle
from an AGN instead of the starburst, as suggested by some optical (Veilleux
\et \cite{Veil95}), NIR (Veilleux, Sanders \& Kim, \cite{veil99}) and ISO (
Tran \et \cite{Tran2001}) spectroscopic studies of ULIRGs.

To explain the observed large discrepancy, the AGN should provide
a conspicuous fraction of the  IR luminosity, generally exceeding 80\%.

As already anticipated we have possibly detected broad emission line components
and [SiVI] emission only in IRAS 00188-0856 and IRAS 00582-0258.
 
It is interesting to note that, in the  IR colour-luminosity diagram (Neff \&
Hutchings \cite{neff}), both IRAS 00188-0856 with 
log(S$_{25\mu m}$/S$_{60\mu m}$)=$-$0.63 
and log(L $_{60\mu m}$/L$_{\odot}$)=12.05, 
and IRAS 00582-0258 with log(S
$_{25\mu m}$/S$_{60\mu m}$)=$-$ 0.55 
and log(L $_{60\mu m}$/L$_{\odot}$)=11.24, fall
in the region of overlap between QSO and Seyfert 1 galaxies.

The presence of a broad component (FWHM$>$2000 km/s) in the \pa\ line, the
possible detection of the high excitation [SiVI] line, and their position in the
IR colour-luminosity diagram, favour the existence of an obscured AGN in IRAS
00188-0856 and IRAS 00582-0258.

\subsubsection{Other diagnostics for unveiling the AGN}
\label{other}

Hints on the nature of the IR luminosity may come also from other wavelength
observations.

In star forming galaxies, FIR and radio emissions are tightly correlated over a wide
range of IR luminosities. Sanders \& Mirabel (\cite{sand1}) report:

\begin{equation}
q=\log \frac{\mbox{F}_{FIR}/(3.75\times 10^{12}\mbox{Hz})}{F_\nu
(1.49\mbox{GHz})/(\mbox{W m}^{-2}\mbox{Hz}^{-1})}\simeq 2.35\pm 0.2
\label{qeq}
\end{equation}

\noindent where F$_{FIR}$=1.26$\times$10$^{-14}$(2.58$S_{60\mu m}$ + $S_{100\mu m}$) $\mbox{Wm}^{-2}$, with $S_{60}$ and $S_{100}$
expressed in Jy.
Radio observations of all our target objects have been reported in Table
\ref{irprop}. After extrapolating a few 843MHz data to 1.4GHz by assuming
a radio slope F$_{\nu}$~$\propto$~$\nu^{-0.8}$, we have evaluated the q parameter,
reported in Table \ref{irprop}.

All the objects, including 
IRAS00188-0856 and IRAS00582-0258,
but IRAS22491-1808 and IRAS23389-6139, fall on top of the 
FIR-radio relation of starburst galaxies. 
Panuzzo \et (\cite{pan03}) have shown that in normal star-forming galaxies the
radio luminosity is an excellent indicator of star formation and that deviations
from this relation are due to variations of the corresponding FIR luminosity as
the latter depend more on the details of the obscuration. 
In contrast, in a very young obscured starburst there is an
excess of FIR emission because core collapsed supernovae, thought to provide the
90\% of radio emission, have a delay of a few million years. For the same
reason, if the starburst is in a late phase and the star formation decreased
exponentially, there is an excess of radio emission (Bressan, Silva \& Granato
\cite{bress01}). 

Based on different arguments, Smith \et
(\cite{sm98}), Farrah \et (\cite{Farrah}) and Prouton \et (\cite{prout04}), have
shown that the contribution of the putative AGN to the radio luminosity in
obscured starbursts (not harbouring radio-loud sources) is low.
Thus a significant contribution to the IR from the AGN would significantly raise
the value of q above the FIR-radio correlation. This might be particularly relevant
for IRAS00188-0856 and IRAS 00582-0258 where, as we have already seen, there are
hints for the presence of the AGN from our own observations, but in fact these
galaxies show a q parameter typical of starbursts indicating
that the contribution of the AGN in the IR is not high.

Another diagnostic for the presence of the AGN is a low value of the ratio of the
line to continuum  emission of the 7.7 $\mu$m PAH feature, as measured by ISO
spectroscopy (Lutz \et \cite{Lutz98}). Unfortunately, not all our galaxies have
this ratio measured. Following this indicator (Table \ref{irprop}), three sources
could harbour an obscured AGN, IRAS 00188-0856, IRAS 23230-6926 and IRAS 23389-
6139. However only the first source shows evidence for an AGN from our NIR
observations. This may give further support to criticism  on the use of this
indicator recently raised by Farrah \et (\cite{Farrah}) (but see also Prouton \et
\cite{prout04}).

There are also XMM-Newton hard X-ray observations for IRAS 20100-4156 
(Franceschini \et \cite{frabra}). IRAS 20100-4156
shows a faint X-ray flux of S(2-10 keV)$\simeq$ 1.9$\times$ 10$^{-14}$
erg s$^{-1}$cm$^{-2}$, with a (2-10Kev)/(8-1000$\mu$m) flux ratio $\simeq$4.75
$\times$ 10$^{-5}$. Unfortunately, its large distance and consequent poor X-ray
photon statistics prevented any careful X-ray spectral analyses. Although the
total source flux is slightly above that expected from a starburst of similar FIR
power (Franceschini \et \cite{frabra}), there is no definite evidence, from hard
X-rays, that the galaxy hosts an absorbed AGN. For instance MKN231, a typical Sy1
galaxy, is about 34 times brighter, between 2 and 10 keV, than IRAS 20100-4156,
in spite of having about the same IR luminosity. 

In summary, from our own observations and other collected data in the literature we
may conclude that IRAS 00188-0856 and IRAS 00582-0258 could
harbour an AGN. However in all cases the contribution of the AGN to the IR seems
not at the level required to explain the discrepancy between line flux and FIR
emission. 
Indeed the first evidence of a buried AGN in the ultraluminous
galaxy UGC 5101, provided by SPITZER,
allows a contribution to the total luminosity of only  $\leq$10\%
(Armus \et \cite{armus04}).

\subsection{High NIR extinction}

\label{nir-ext}

Another alternative to explain the photon recombination deficit is simply that the
extinction is very high, even at NIR wavelengths, and that what we see from optical lines
is only the external skin of the starburst.

To evaluate the possible range of that kind of extinction we have compared the observed
values of the \pa/IR (or \bg/IR) ratio with that predicted by models of normal star-
forming galaxies. In this way we will obtain the excess extinction over that
of a normal star forming galaxy.
From  Panuzzo \et (\cite{pan03}) we get the following average relations
between line and IR (8\mum -- 1000\mum) emission:

S(\ha)/S(IR)=6.56$\times$10$^{-3}$,

S(\pa)/S(IR)=7.9$\times$10$^{-4}$, and

S(\bg)/S(IR)=6.32$\times$10$^{-5}$

The observed values  of S(\pa)/S(IR) or S(\bg)/S(IR) are reported in Table
\ref{ratio}. The observed intensities of the emission lines have been corrected for
slit losses by multiplying by the factor C$_{IR}$ of Table \ref{losses}.
The average value of the observed S(\pa)/S(IR) ratios is $\simeq$10\%
of that predicted by the models while, in the case of \bg\ line, 
it is $\simeq$40\% of the predicted value.

Assuming that the difference with respect to the models is entirely
due to an extinction larger than in the case of a normal star forming galaxy, we have
derived the corresponding attenuation in magnitudes, in the line and in the visual.
The values of A$_V$ must be considered as in excess of that of normal star forming
galaxies which is typically A$_V\simeq$1 mag. In this way we obtain an average
attenuation at \pa\ of about 2.9 mags, corresponding to an average visual attenuation of
about 20 mags. These values are much higher than those derived from optical emission
lines (cfr. Table \ref{ext}).
Values of A$_V\sim$5-50 have been already found by
Genzel at al. (1998)
from mid-infrared spectroscopy of a  small
sample of ULIRGs and are 
confirmed by recent SPITZER observations of selected ULIRGs (
Armus \et \cite{armus04}).
High optical depths ($\geq$20) at 1\mum~ 
have been also inferred from fitting the far infrared to radio SEDs 
of a sample of compact ULIRGs (Prouton \et \cite{prout04}). 

\begin{table}[!ht]
\begin{center}
\caption{Extinction derived from comparison of observed and predicted
line flux to IR ratios}
\label{ratio}
\begin{tabular}{lrrrr}\hline \hline
IRAS name & Line/IR& Obs/Model$^a$& A(line) & A$_V^b$ \\
\hline
&\multicolumn{3}{c}{\pa}&\\
00188-0856  &  2.73E-05  &  3.45E-02 &3.65  & 24.9\\
00582-0258  &  2.13E-05  &  2.70E-02 &3.92  & 26.7\\
02411+0354  &  1.42E-04  &  1.80E-01 &1.86  & 12.7\\
06206-6315  &  3.97E-05  &  5.03E-02 &3.25  & 22.1\\
19335-3632  &  1.72E-04  &  2.18E-01 &1.65  & 11.2\\
20100-4156  &  4.01E-05  &  5.08E-02 &3.23  & 22.0\\
22206-2715  &  6.53E-05  &  8.27E-02 &2.71  & 18.4\\
22491-1808  &  5.73E-05  &  7.25E-02 &2.85  & 19.4\\
23230-6926  &  7.86E-05  &  9.95E-02 &2.51  & 17.0\\
23389-6139  &  4.56E-05  &  5.78E-02 &3.10  & 21.1\\
\hline                                        
&\multicolumn{3}{c}{\bg}&\\
00085-1223  &  1.88E-05  &  2.98E-01 & 1.31 & 13.5  \\
01077-1707  &  3.02E-05  &  4.77E-01 & 0.80 & 8.3 \\
23128-5919  &  3.08E-05  &  4.88E-01 & 0.78 & 8.0 \\
\hline
\end{tabular}
\end{center}
\footnotesize
a) Observed L(\pa)/IR = 7.9E-4 and L(\bg)/IR = 6.3E-5,
for a normal star forming galaxy (Panuzzo et al. 2003)\\
b) A$_{Pa\alpha}$/A$_V$=0.147 and A$_{Br\gamma}$/A$_V$=0.097 from
Calzetti et al. (2000)
\end{table}

We also notice that, in the case of \pa\ galaxies,
even assuming an AGN contribution to the IR
of the order of
50\% (which is large, following Prouton \et \cite{prout04}),
the visual attenuation would still remain between 15 and 25 mags.

As a second point, we notice that 
in the galaxies observed in the Br$\gamma$ domain
the extinction is lower
than that obtained for the sample with \pa\, in agreement with
a slab attenuated model.
In fact, for such a model we expect A(\bg)$\simeq$0.66
$\times$A(\pa) (Calzetti et al. 2000).
The average extinction of the \bg\ sample is A(\bg)$\simeq$0.96, 
while taht of the \pa\ sample is A(\pa)$\simeq$2.9, with a ratio of
0.33. Though we are comparing different galaxies and the statistics is low 
(but notice
that our \bg\ galaxies lie on the relation defined by the more exhaustive sample of
Goldader et al. 1997), this may indicate that the photon deficit is {\it wavelength
dependent}, an thus favour the effects of a complex extinction
geometry over those of 
absorption by dust within HII regions, discussed below.

\subsection{Dust within the HII regions}

\label{dust}

In presence of dust within the ionized regions
of the starburst, only a fraction
$f$ of Ly-continuum photons may effectively ionize the gas, while the
remaining (1-$f$) is absorbed. 
This effect causes a deficit of recombination photons, with respect to
the FIR emission.
To explain the 
observed discrepancy between NIR line emission and IR flux, 
dust within HII regions should absorb 
about 80\% of the ionizing flux.

Dust absorption within HII regions has been 
invoked by Hirashita \et (\cite{Hira}) 
to explain the anomalous 
\ha\ to UV flux, observed in a  sample of IUE selected star forming galaxies. 
Hirashita \et (\cite{Hira}) have estimated an
average value of (1-$f$)$\simeq$50\%,
with some objects reaching (1-$f$) $\simeq$80\%. 
However, it is worth
recalling that the conclusions reached by Hirashita \et (\cite{Hira}) are
based on a single screen extinction model. By
analyzing the same data, Panuzzo \et (\cite{pan03}) have instead found that
this effect may result from ``age-selective'' extinction, namely that
younger populations are more extinguished than older ones, the latter still
contributing to the UV flux but not to nebular emission.

Dust within ionized regions has been indicated
also by Luhman et al. (2003) as a possibility
to explain the [CII] deficit relative to FIR, observed in their sample of
ULIRGs. In fact dust would not only affect
the pool of ionizing photons, but  would also 
inhibit the penetration
of 13.6eV$-$6eV photons  
(thought to be responsible of 
the [CII] excitation) in the photo-dissociation regions, 
while preserving the overall FIR emission.

If confirmed by further studies, a 80\% depression of the ionizing flux
would deeply challenge any determination of SFR from even the less 
extinguished emission lines.

\subsection{Age of the starbursts}

\label{ages}

Estimators of the SFR are usually derived assuming 
a continuous star formation rate. 
This is essentially correct when dealing with normal galaxies, but
starburst galaxies are, by definition, currently dominated by a single episode,
with a duration which is generally comparable to the characteristic times 
of the star formation indicators.  
In such circumstances the use of different calibrators may lead to
significant discrepancies. For instance, this may be the case of the Radio and IR SFR
indicators, as discussed by Bressan, Silva \& Granato (\cite{bress02}).

To get hints on the evolutionary status of our galaxies we have analyzed
the equivalent widths (EW) of hydrogen emission lines 
(see e.g. Terlevich et al. 2004).
These indicators  are relatively unaffected by dust extinction in at least two 
cases, namely when both continuum and line emission 
arise from the same population,
or when attenuation can be approximated
by an inter·vening absorbing slab. In the latter case
however, the equivalent width may be affected by the presence of the
old population, outside the starburst region.
\begin{figure}[!ht]
\centering
\includegraphics[width=0.48\textwidth]{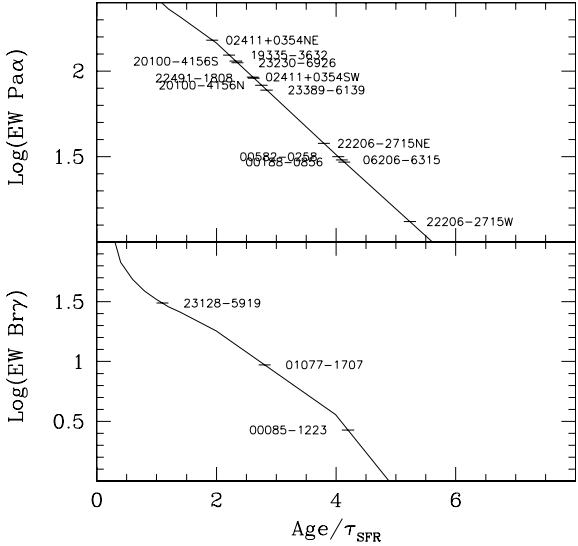}
\caption{Starburst ages derived from the equivalent width of \pa\ and \bg\
emission. The SFR of the  models decreases exponentially with e-folding
time of $\tau_{SFR}$=25Myr. 
The EW of the models is plotted against the ratio between 
the current age and the SFR e-folding
time so that the figure does not change significantly by considering other 
plausible values of  $\tau_{SFR}$.   
Observed values are plotted as horizontal segments.
}
\label{ewpa}
\end{figure}
Figure \ref{ewpa} compares the observed  \pa\ and \bg\ 
equivalent widths of our galaxies 
with those predicted by starburst models. 
The models have been constructed by adding nebular emission to the spectral energy
distribution of simple stellar populations of solar metallicity, as described in
Bressan, Poggianti \& Franceschini (\cite{bress01}). 
The adopted star formation e-folding time is
$\tau_{SFR}$=25, but since we are interested in the ratio between
the age of the starburst and the e-folding time, 
our conclusions do not change significantly
by assuming other plausible values for $\tau_{SFR}$.

From Figure \ref{ewpa}, we observe a lack of young objects, with only one source,
IRAS 23128-5919, appearing younger 
than twice the SFR e-folding time. The other galaxies have ages that are either
between two and three times the SFR e-folding time, or
around four times the e-folding time.
Taken at face value, these estimates would place  IRAS 00188-0856,
IRAS 00582-0258, IRAS 06206-6315 and  IRAS 22206-2715NE
in a {\sl post-starburst} phase.
Part of the photon deficit discrepancy could then originate simply from
our use of stationary models. However,
given the high current IR luminosity of the latter galaxies,
their bolometric luminosity at the peak SFR would have 
been unreasonably large, 
between 3$\times$10$^{13}$L$_\odot$ and 2$\times$10$^{14}$L$_\odot$.

It is worth noticing that, being the average distance
of the \pa\ sample about four times that of the \bg\ sample,
our slit is sampling an intrinsic galaxy area which increases by 
at least one order of magnitude going from the \bg\ sample to the
\pa\ sample. In spite of that, both  
the equivalent width of the lines and the corresponding SFR
show a comparable distribution in the two samples (though the \bg\ sample is
not statistically significant).
This not only suggests that the line emission originates from
the central regions, but also indicates 
that the equivalent widths are dominated by 
sources that share the same spatial distribution.
For instance, they should not be significantly affected by an underlying
old population, as expected in 
vigorous starbursts (Mayya et al. \cite{mayya}).

We conclude that the observed distribution of EWs, skewed toward old ages,
is actually another aspect of the photon deficit
problem, {\sl independent from bolometric luminosity
considerations.}
Unfortunately a statistical argument based on the 
distribution of EWs cannot help to
discriminate between  the different causes.
In fact, strong age selective extinction could 
make the starburst to appear older than it is;
absorption of ionizing photons
could diminish the equivalent width, because
the K band continuum of the ionizing cluster
would suffer much less attenuation;
and finally the K-band continuum could be 
dominated by the presence of an AGN, still compatible with 
the FIR/Radio correlation (Prouton et al. 2004).

\section{Conclusions and Perspectives}

\label{conc}

We have obtained NIR medium dispersion long slit spectroscopy with SOFI at NTT of
ten luminous infrared galaxies having suitable redshift to push their \pa\
emission in the Ks band. We included also three objects with lower redshift, for
which the \bg\ line was in the Ks band.

We have found that the \pa\ emission,
even corrected for slit losses and for extinction
estimated by comparing our data with optical spectroscopy, 
is significantly less than that expected from a starburst of 
corresponding bolometric luminosity.
The discrepancy is lower for the galaxies observed in the
\bg\ region, but in general we
confirm the existence of a deficit of recombination photons
first pointed out in ULIRGs by Goldader \et (\cite{gold95}).

Furthermore, 
in  IRAS 00188-0856 and
IRAS 00582-0258 we find evidence for significant broadening of the \pa\ line 
and for the presence of [SiVI] coronal line.
However these two galaxies fall on top of the FIR/Radio correlation,
indicating that, though present, the AGN does not
dominate the far infrared emission.
For all other sources we do not find  evidence of 
AGN contribution and we argue that the studied galaxies appear to be 
predominantly powered by a nuclear starburst.

Based on current data alone it is impossible to disclose the origin of the 
recombination photon deficit, and we may only advance the following hints.

The galaxies may harbour a highly attenuated star forming
region. In this case an estimate of the attenuation can be obtained 
from the comparison the observed line/IR emission ratio
with the predictions of models for normal star-forming galaxies.
Then the average attenuation would be A$_V\simeq$10-25mag and
A$_{Pa\alpha}\simeq$2-4mag. These figures are slightly lower when derived from
the \bg/IR ratio (A $_{Br\gamma}\simeq$1mags) and are consistent with the
decrement of attenuation going from \pa\ to the \bg\ wavelength region. A
significant attenuation of the nuclear star forming region even in the NIR comes
out to be in agreement with the large molecular cloud optical depths (
$\tau_{{1\mu}m}\geq$ 20), derived by Prouton \et (\cite{prout04}) from SED
fitting of compact ULIRGs. It is also in agreement with recent SPITZER observations of
ultra luminous galaxies with  buried AGN (Armus et al. 2004).
Assuming that the AGN may contribute about 50\% of the IR flux (a
quite extreme figure, Prouton \et \cite{prout04}), 
would imply only a slightly smaller
attenuation.

Our finding is also compatible with
a scenario where a large fraction ($\simeq$80\%) of the ionizing flux
is absorbed by dust {\it within} the HII regions. 
The required fraction is large but comparable with
some extreme values found by Hirashita \et (\cite{Hira}) in 
their analysis of a sample of UV selected starbursts.
{This effect has been recently invoked as one of the possible 
causes of the [CII] emission deficit relative to FIR observed in
some ULIRGs (Luhman et al. 2003 )}.
Based on MIR to radio SED fitting, Prouton \et (\cite{prout04}) have estimated that
the dust sublimation radius within molecular clouds in compact ULIRGs, is generally a
fraction of a parsec.
Dust may thus survive even in the innermost regions of an HII region
and,  if  such a strong effect is confirmed, it will challenge our
ability to derive SFR and/or to study inner environmental conditions,
from even the most un-extinguished emission lines.

Though the above alternative scenarios are both compatible with the present
data, their predictions at longer wavelengths are markedly different and suggest
that a decisive test to disentangle between high nuclear obscuration and dust
absorption within HII regions, would be to look at the Br$\alpha$ 4.05 $\mu$m line
emission.
In fact, assuming the Calzetti et al. (2000) extinction law and an intrinsic ratio L(
Pa$\alpha$)/L(Br$\alpha$)$\simeq$3.9 (Panuzzo \et \cite{pan03}) we get
\begin{displaymath}
L(Br\alpha)/L(Pa\alpha) \simeq 0.26 \times 10^{0.04 A_V}.
\end{displaymath}
Thus if the \pa\ deficit is due to absorption of ionizing photons by 
dust within HII region and A$_V$$\simeq$3 mag, 
as derived from optical and NIR-optical emission
line ratios, then the expected ratio L(Br$\alpha$)/L(Pa$\alpha$) is
$\simeq$0.34. On the other hand, if the \pa\ emission comes from a dust
enshrouded region with say A$_V$$\simeq$20 mag, then L(Br$\alpha$)/L(Pa$\alpha$)
$\simeq$1.6, a factor about five times larger than in the previous case.

\begin{acknowledgements}
We tank the anonymous referee for her/his comments and suggestions and
M. Clemens for discussions and careful reading of the manuscript.
A.B. acknowledges warm hospitality by INAOE and
J.R.V. acknowledges warm hospitality by INAF, Osservatorio Astronomico di Padova.
S.B. acknowledges support by
ASI research grant no. I/R/062/02.
This research was partially supported by the European
Commission Research Training Network `POE' under contract
HPRN-CT-2000-00138 and by MURST under COFIN n. 2001/021149 
\end{acknowledgements}


\begin{thebibliography}{}

\bibitem[2004]{armus04} Armus, L.; \et\ 2004, AAS, 204, 3319
\bibitem[2000]{bar} Barger, A. J., Cowie, L. L. \& Richards, E. A. 2000, AJ, 119, 209.
\bibitem[2003]{ste} Berta, S., Fritz, J., Franceschini, A., Bressan, A. \& Pernechele, C.: 2003, \aap, 403, 119.
\bibitem[2002]{bress02} Bressan, A., Silva, L., \& Granato, G.~L.\ 2002, \aap, 392, 377
\bibitem[2001]{bress01} Bressan, A., Poggianti, B., \& Franceschini, A.\ 2001, QSO Hosts and Their Environments, 171
\bibitem[1994]{bress94} Bressan, A., Chiosi, C., Fagotto, F.\ 1994, ApJS, 94, 63 
\bibitem[2002]{bush} Bushouse, H.A., Borne, K.D., Colina, L., Lucas, R.A., Rowan-Robinson, M., Baker, A.C., Clements, D.L., Lawrence, A., Oliver, S., 2002, ApJSS, 138, 1
\bibitem[2000]{cal} Calzetti, D., Armus, L., Bohlin, R.~C., Kinney, A.~L., Koornneef, J., \& Storchi-Bergmann, T.\ 2000, \apj, 533, 682
\bibitem[1996]{condon} Condon, J.~J., Helou, G., Sanders, D.~B., \& Soifer, B.~T.\ 1996, \apjs, 103, 81
\bibitem[1998]{con98} Condon, J. J.; Cotton, W. D.; Greisen, E. W.; Yin, Q. F.; Perley, R. A.; Taylor, G. B.; Broderick, J. J., 1998, AJ, 115, 1693
\bibitem[1997]{duc} Duc, P.-A., Mirabel, I. F. \& Maza, J. 1997, A\&AS, 124, 533
\bibitem[1999]{elb} Elbaz, D., Cesarsky, C.J., Fadda, D., \et~1999, A\&A, 351, 37
\bibitem[2003]{Farrah} Farrah, D., Afonso, V., Efstathiou, A., Rowan-Robinson, M. ,Fox,  M., Clements, D.,\ 2003, MNRAS, 343, 585
\bibitem[2003]{Ferland} Ferland, G.~J.\ 2003, \araa, 41, 517 
\bibitem[1995]{fish} Fisher, K.B., \et, 1995, ApJS, 100, 69
\bibitem[2001]{fran01} Franceschini A., Aussel H., Cesarsky C., Elbaz D., Fadda D.: {\em A\&A}, 2001, 378, 1.
\bibitem[2003]{frabra} Franceschini, A., Braito, V.,Persic, M. et al. 2003, MNRAS, 343, 1181
\bibitem[2000]{gear2000} Gear, W.~K., Lilly, S.~J., Stevens, J.~A., Clements, D.~L., Webb, T.~M., Eales, S.~A., \& Dunne, L.\ 2000, \mnras, 316, L51
\bibitem[1998]{gen} Genzel, R., Lutz, D., Sturm, E., \et~1998, ApJ, 498, 579
\bibitem[1997]{gold97} Goldader, J.~D.; Joseph, R.~D.; Doyon, R.; Sanders, D.~B.\ 1997, ApJS, 108, 449  
\bibitem[1995]{gold95} Goldader, J.~D., Joseph, R.~D., Doyon, R.; Sanders, D.~B.\ 1995, \apj,
444, 97
\bibitem[2000]{gran00} Granato, G.L., Lacey, C.G., Silva, L., Bressan, A., Baugh, C.M., Cole, S., Frenk, C.S.\ 2000, \apj, 542, 710
\bibitem[2003]{Hira} Hirashita, H., Buat, V., Inoue, A. K., 2003, A\&A, 410, 83
\bibitem[1987]{humm87} Hummer, D. G.; Storey, P. J.\ 1987, MNRAS, 224, 801
\bibitem[2002]{ivis2002} Ivison, R.~J.~et al.\ 2002, \mnras, 337, 1
\bibitem[1998]{kenn} Kennicutt, R.C. 1998, ARAA, 36, 189
\bibitem[1997]{kim} Kim, A. G., Gabi, S., Goldhaber, G., \et~1997, ApJ, 476, L63
\bibitem[2003]{luhman} Luhman, M.L., \et, 2003, ApJ, 594, 758
\bibitem[1998]{Lutz98} Lutz, D., Spoon, H.~W.~W., Rigopoulou, D., Moorwood, A.~F.~M., \& Genzel, R.\ 1998, \apjl, 505, L103
\bibitem[2003]{mauch03}Mauch, T.; Murphy, T.; Buttery, H. J.; Curran, J.; Hunstead, R. W.;
Piestrzynski, B.; Robertson, J. G.; Sadler, E. M., 2003, MNRAS, 342, 1117
\bibitem[2004]{mayya} Mayya, Y.D., Bressan, A., Rodriguez, M., Valdes, J.R., Chavez, M., 2004, ApJ, 600, 188
\bibitem[1998]{moor} Moorwood A., Cuby J.G. \& Lidman C. 1998, The Messenger 91, 9
\bibitem[1992]{neff} Neff, S. G., Hutchings, J. B., 1992, AJ, 103, 1746
\bibitem[2003]{pan03} Panuzzo, P., Bressan, A., Granato, G.~L., Silva, L., \& Danese, L.\ 2003, \aap, 409, 99
\bibitem[2003]{pernec} Pernechele, C., Berta, S., Marconi, A., Bonoli, C., Bressan, A., Franceschini, A., Fritz, J., \& Giro, E.\ 2003, MNRAS, 338, L13
\bibitem[1998]{pick} Pickles, A.J. 1998, PASP, 110, 863
\bibitem[2001]{pogg} Poggianti, B.\,M., Bressan, A., \& Franceschini, A., 2001, ApJ, 550, 195
\bibitem[2004]{prout04} Prouton, O. R.; Bressan, A.; Clemens, M.; Franceschini, A.; Granato, G. L.; Silva, L.\ 2004, A\&A, 421, 115
\bibitem[1999]{rig1} Rigopoulou, D., Spoon, H.~W.~W., Genzel, R., Lutz, D., Moorwood, A.~F.~M., \& Tran, Q.~D.\ 1999, \aj, 118, 2625
\bibitem[1996]{sand1} Sanders, D. B. \& Mirabel, I.F. 1996, ARAA, 34, 7479
\bibitem[1998]{silva98} Silva, L., Granto, G.L., Bressan, A., Danese, L.\ 1998, \apj, 509, 103
\bibitem[2000]{sma} Smail, I., Ivison, R.J., Owen, F.N., \et~2000, ApJ,
528, 612
\bibitem[2003]{smaetal2003} Smail, I., Ivison, R.~J., Gilbank, D.~G., Dunlop, J.~S., Keel, W.~C., Motohara, K., \& Stevens, J.~A.\ 2003, \apj, 583, 551
\bibitem[1998]{sm98} Smith, H.E.; Lonsdale, C.J.; Lonsdale, C.J.\ 1998, A\&A, 492,137
\bibitem[2003]{Stevens} Stevens, J.~A.~et al.\ 2003, NATURE, 425, 264
\bibitem[2001]{Tran2001} Tran Q. D., et al, 2001, ApJ, 552, 527
\bibitem[2004]{Terlevich} Terlevich, R., Silich, S., 
Rosa-Gonz{\' a}lez, D., \& Terlevich, E.\ 2004, \mnras, 348, 1191 
\bibitem[2002]{vanzi} Vanzi, L., Bagnulo, S., Le Floc'h, E., Maiolino, R., Pompei, E., Walsh, W., 2002, A\&A, 386, 464
\bibitem[1999]{veil99} Veilleux, S., Kim, D.-C., \& Sanders, D.~B., 1999, ApJ, 522, 113
\bibitem[1999]{veil99a} Veilleux, S., Sanders, D.~B., \& Kim, D.-C., 1999, ApJ, 522, 139
\bibitem[1995]{Veil95} Veilleux S., Kim, D.-C., Sanders D. B., Mazzarella J. M., Soifer B. T., 1995, ApJS, 98, 171
\bibitem[1993]{zen} Zenner, S., Lenzen, R., 1993, A\&ASS, 101, 363
\end{thebibliography}
\end{document}